\documentclass[aps, prb, twocolumn, showpacs, 10pt, floatfix, superscriptaddress, floatfix, longbibliography, nofootinbib]{revtex4-2}

\usepackage{setspace}

\usepackage[english]{babel}
\usepackage[utf8]{inputenc}
\usepackage[T1]{fontenc}

\usepackage{amsmath}
\usepackage{braket}
\usepackage{amssymb}
\usepackage[normalem]{ulem}
\usepackage{dsfont}
\usepackage{bm}
\usepackage[dvipsnames]{xcolor}
\usepackage{stmaryrd}
\usepackage{bbm}
\usepackage{mathtools}

\usepackage{multirow}
\usepackage{enumitem}

\usepackage{color}
\usepackage[dvipsnames]{xcolor}

\usepackage[most]{tcolorbox}
\newtcbox{\othermathbox}[1][]{nobeforeafter, math upper, tcbox raise base, enhanced, rounded corners, colback=black!5, colframe=black, left=0.3em, top=0.3em, right=0.3em, bottom=0.4em}
\usepackage{empheq}
\usepackage[colorlinks,allcolors=blue]{hyperref}

\usepackage{enumerate}

\newcommand{\be}{\begin{equation}}
\newcommand{\ee}{\end{equation}}

\newcommand{\Ad}{\text{Ad}}
\newcommand{\ad}{\text{ad}}
\newcommand{\dd}{\text{d}}
\newcommand{\der}{\partial}

\newcommand{\ds}{\displaystyle}
\newcommand{\eg}{\textit{e.g.}\ }

\newcommand{\ie}{\textit{i.e.}\ }

\newcommand{\cA}{{\cal A}}

\newcommand{\cH}{{\cal H}}

\newcommand{\cM}{{\cal M}}
\newcommand{\cO}{{\cal O}}

\newcommand{\mg}{\mathfrak{g}}

\newcommand{\sfC}{\mathsf{C}}

\newcommand{\sfK}{\mathsf{K}}

\newcommand{\sfs}{\mathsf{s}}
\newcommand{\sfS}{\mathsf{S}}

\newcommand{\II}{\mathbb{I}}

\newcommand{\RR}{\mathbb{R}}

\newcommand{\ZZ}{\mathbb{Z}}

\renewcommand\i[1]{\textit{#1}}

\usepackage{changepage}

\begin{document}

\title{Berry Phases in the Bosonization of Nonlinear Edge Modes}

\author{Mathieu Beauvillain}
\email{mathieu.beauvillain@polytechnique.edu}
\affiliation{CPHT, CNRS, \'Ecole polytechnique, Institut Polytechnique de Paris, 91120 Palaiseau, France}

\author{Blagoje Oblak}
\email{oblak@math.univ-lyon1.fr}
\affiliation{Universit\'e Claude Bernard Lyon 1, ICJ UMR 5208, CNRS, 69622 Villeurbanne, France}

\author{Marios Petropoulos}
\email{marios.petropoulos@polytechnique.edu}
\affiliation{CPHT, CNRS, \'Ecole polytechnique, Institut Polytechnique de Paris, 91120 Palaiseau, France}

\begin{abstract}
We consider chiral, generally nonlinear density waves in one dimension, modelling the bosonized edge modes of a two-dimensional fermionic topological insulator. Using the coincidence between bosonization and Lie-Poisson dynamics on an affine U(1) group, we show that wave profiles which are periodic in time produce Berry phases accumulated by the underlying fermionic field. These phases can be evaluated in closed form for any Hamiltonian, and they serve as a diagnostic of nonlinearity. As an explicit example, we discuss the Korteweg-de Vries equation, viewed as a model of nonlinear quantum Hall edge modes.
\end{abstract}

\maketitle

%


\section{Introduction}
\label{secIN}

Geometric observables, and geometric phases in particular, are ubiquitous in both classical and quantum physics \cite{Berry,BerryClassical}. In the solid-state context, they play a key role in the quantum geometry of Brillouin zones, determining \eg quantized response coefficients of topological insulators \cite{TKNN,BernevigBook}. In hydrodynamics, similar geometric objects contribute to the robustness of certain equatorial waves, which may be viewed as edge modes of an `atmospheric topological insulator' \cite{Delplace_2017}.  All these cases involve Berry curvatures with nontrivial Chern classes, but it is worth stressing at the outset that even topologically \i{trivial} Berry phases matter for physics: illuminating examples are provided by the Thomas precession of relativistic electrons \cite{Thomas,Oblak:2017oge,Oblak:2017ptc}, Aharonov-Bohm phases\footnote{The standard presentation of the Aharonov-Bohm effect involves a solenoid that makes the fundamental group of space nontrivial, but this is an unnecessary complication. Indeed, Aharonov-Bohm phases also occur for adiabatic transport of charged particles in a magnetic field: the latter plays the role of a Berry curvature, and parameter space is space itself---say the plane $\RR^2$---, with trivial topology.} \cite{Aharonov}, and the odd viscosity of quantum Hall systems \cite{AvronSeiler,Levay}. In short, even local quantum geometry (as opposed to its global topological consequences) affects natural phenomena.

The present work is devoted to (topologically trivial) Berry phases produced by bosonic edge modes on the boundary of any two-dimensional (2D) topological insulator. Such edge excitations are gapless and chiral, and described by an effective 1D conformal field theory at leading order in the thermodynamic limit \cite{Halperin,Hatsugai,WenChiral,WenHall}. However, finite-size subleading corrections generally exist; they give rise to nonlinear and/or dispersive edge dynamics, typically observable in numerics \cite{NardinCarusotto,Nardin:2022mto,Nardin_2023,Nardin:2024dyk} and quantum simulators \cite{Leykam,Leonard:2022ndq,Binanti:2023ozm}. The prototypical setup exhibiting such a behaviour is the quantum Hall effect \cite{vonKlitzing,Tsui} with its edge magnetoplasmons, whose nonlinearity has been studied for decades \cite{Agam:2001eu,Abanov:2005nt,Bettelheim:2006un,Wiegmann:2011np} and was recently revisited in the aforementioned numerical works \cite{NardinCarusotto,Nardin:2022mto,Nardin_2023,Nardin:2024dyk}. We will see here that the Berry phases produced by periodic edge modes act as a diagnostic of nonlinearity, in the sense that they are `trivial' for free theories but nontrivial for dispersive and/or nonlinear systems. As a bonus, these phases follow from \i{classical} symplectic geometry and thus require no assumption of adiabaticity. Their future measurement therefore seems possible, for example in electronic interferometers that have recently been used to observe anyon statistics \cite{Bartolomei:2020qfr,Feldman,Frigerio}.

In practice, our derivation rests on the fact that bosonization \cite{Sugawara:1967rw,Sommerfield:1968pk,Haldane:1981zz,Haldane:1981zza,Giamarchi} can be reformulated in terms inspired by geometric hydrodynamics \cite{Karabali:1991hm,Iso:1992aa,Das:1991qb,Das:1991uta,Sakita:1996ne,Delacretaz:2022ocm} (see also \cite{Huang:2023hbt,Huang:2024uap}). The latter dates back to Arnold's seminal work on incompressible fluids \cite{ArnoldOrigin}, and roughly consists in viewing fluid flows as paths in an infinite-dimensional space of diffeomorphisms (smooth deformations) of the fluid domain \cite{ArnoldKhesin,Khesin,Modin}. More generally, this approach yields tools to treat field theories as Hamiltonian systems on infinite-dimensional Lie groups or their coadjoint orbits. We shall therefore refer to the evolution equations governing such systems as \i{Lie-Poisson equations}, stressing that they blend two mathematical tools: Lie groups and symplectic (Poisson) geometry \cite{ArnoldKhesin,Khesin}. For instance, it was recently pointed out in \cite{Delacretaz:2022ocm} that Fermi liquid theory is a Lie-Poisson system on a group of canonical transformations.

\begin{figure}[t]
\includegraphics[width=.28\textwidth]{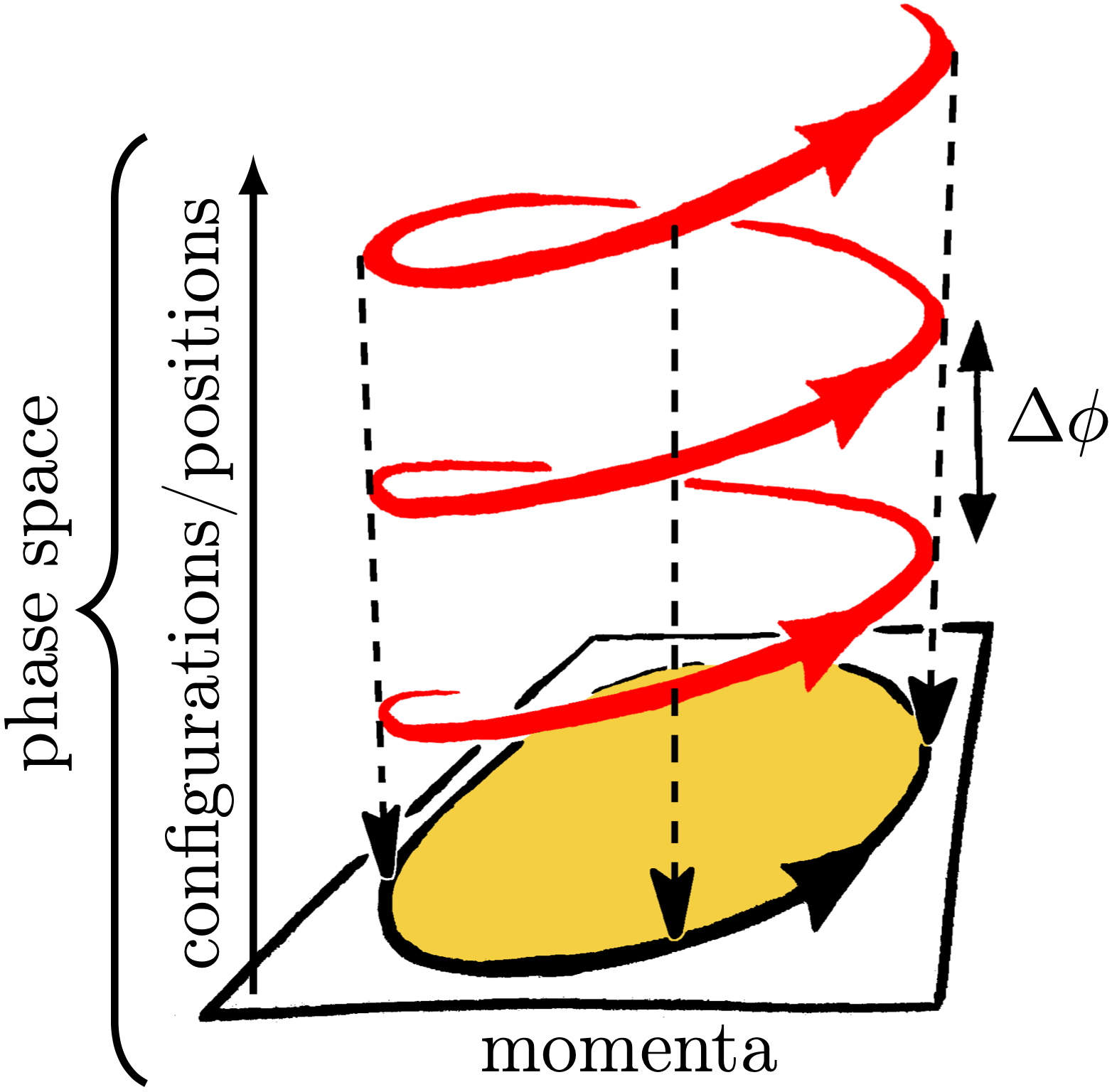}
\caption{A cartoon of Lie-Poisson dynamics, seen as a path in an infinite-dimensional phase space (\eg a space of fluid configurations and velocities, as in hydrodynamics). As the system evolves (in red), it probes the local geometry of phase space. Paths whose momenta are periodic turn out to be sensitive to the (yellow) symplectic area that they enclose on a coadjoint orbit. This area is essentially an Aharonov-Bohm phase, \ie a Berry phase that contributes to the system's change of configuration ($\Delta\phi$) during one period of its momentum. See fig.~\ref{fi4} for a version of the same picture in the case of chiral density waves.}
\label{FIFLOW}
\end{figure}


By construction, Lie-Poisson equations probe the geometry of their configuration space or phase space, whose metric or symplectic form often have physical consequences. An example was put forward in \cite{Oblak:2020jek}, where a drift velocity of fluid parcels was related to certain `symplectic fluxes' on coadjoint orbits of a group of diffeomorphisms \cite{Oblak:2017ect,Oblak:2017ptc}. These fluxes, in turn, could be viewed as Aharonov-Bohm phases \cite{Aharonov} when seeing the symplectic two-form as an effective magnetic field. In the same vein, we show here that the bosonization of chiral density waves is in fact a Lie-Poisson equation on an affine U(1) group.\footnote{In physics, `affine groups' and `current algebras' are more commonly called Kac-Moody groups and algebras. Here we follow instead the convention advocated in \cite{Bardakci:2009wp}, the name `Kac-Moody' being better suited when referring to general algebraic systems, which include affine and hyperbolic algebras and groups.} We then exploit the tools of \cite{Oblak:2020jek} to prove that periodic waves produce Berry phases, again measuring symplectic fluxes on an infinite-dimensional coadjoint orbit; see fig.~\ref{FIFLOW}. The orbit is homotopic to a point, so there is no topological invariant associated with the underlying Berry curvature. Nevertheless, the Berry phases found here very much affect quantities that are expected to be observable: as we show, adding the Berry phase to a suitable dynamical term predicts the overall change of phase of the fermionic field.

The plan of the paper is as follows. In section \ref{seRELP}, we briefly review the bosonization formalism beyond conformal field theory \cite{Karabali:1991hm,Iso:1992aa,Das:1991qb,Das:1991uta,Sakita:1996ne,Delacretaz:2022ocm}, and relate it to Lie-Poisson dynamics on an affine U(1) group. This is used in section \ref{SEBERR} to compute the Berry phase produced by any periodic density wave, and to show that it affects the phase of the parent fermionic field. We conclude in section \ref{seconc} and move mathematical details to appendices \ref{appa} and \ref{appb}, respectively devoted to generalities on Lie-Poisson systems and to the specifics of Berry phases therein.

\section{Bosonization is Lie-Poisson}
\label{seRELP}

This section establishes the general link between bosonization and Lie-Poisson dynamics on an affine U(1) group, for any theory of chiral fermions in 1D with any Hamiltonian. In practice, we begin by reviewing semiclassical bosonization \cite{Karabali:1991hm,Iso:1992aa,Das:1991qb,Das:1991uta,Sakita:1996ne} from the perspective of constrained Hamiltonian systems \cite{HenneauxTeitelboim}, then introduce the affine U(1) group and Lie-Poisson dynamics thereon. The two are related in a self-contained manner, but generalities on Lie-Poisson dynamics are collected in appendix \ref{appa} to streamline the presentation. We refer \eg to \cite[sec.~2]{Oblak:2020jek} for a broader pedagogical introduction whose motivations and language are similar to ours. For general background on Lie groups and symplectic geometry, see \eg \cite{Abraham}.

\subsection{Bosonization in brief}
\label{SEBO}

Here we review the bare bones of bosonization in the classical limit, for any theory of 1D chiral fermions. We start by introducing the fermionic action functional, then discuss its global U(1) symmetry and show that the latter entails a bosonic reformulation of the dynamics. We end by proving that Poisson brackets of the U(1) charge density satisfy a current algebra when restricted to a space of wave profiles with fixed total charge, and compare this to the more standard derivation of the U(1) current algebra from normal-ordered operators. The discussion involves constraints and Dirac brackets, so we refer the unacquainted reader to \cite{HenneauxTeitelboim} for an introduction.

\begin{figure}[t]
\includegraphics[width=.25\textwidth]{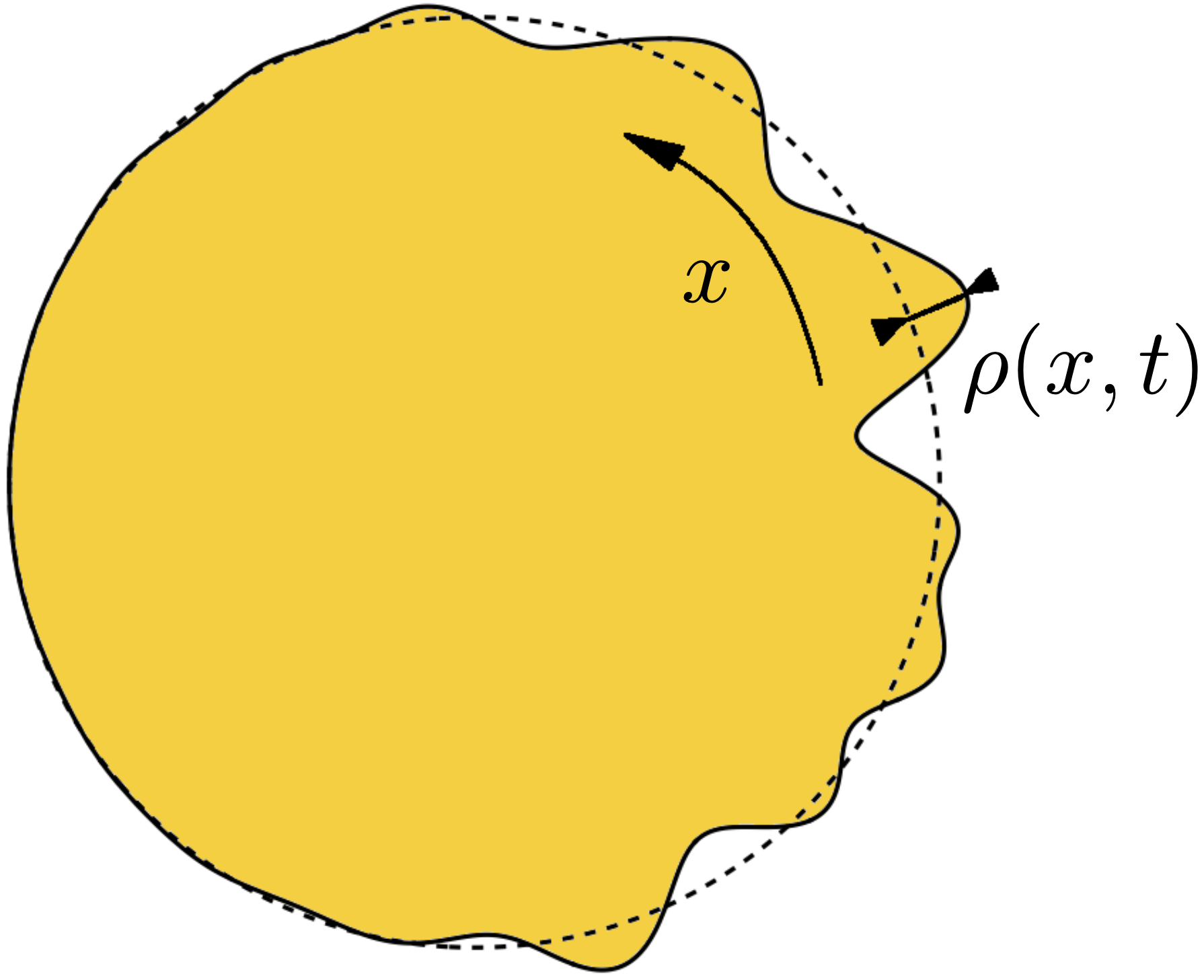}
\caption{A 2D topological insulator, loosely shown as the yellow domain occupied by an incompressible quantum liquid whose edge supports chiral density waves. For comparison, the shape of the ground state is indicated by a dashed line. The edge is assumed to be a circle with angle $x\in[0,2\pi[$, though more complicated shapes can be treated with the same formalism provided the angle coordinate $x$ is such that the leading-order dynamics is nondispersive and free \cite{Oblak:2023bfn}.}
\label{FIWAVE}
\end{figure}

\medskip
\noindent\textbf{Dynamics of chiral fermions.}
Consider a circle whose points are labelled by an angular coordinate $x\sim x + 2\pi$. It may be viewed as the 1D boundary of some 2D topological insulator, as in fig.~\ref{FIWAVE}: think for instance of a quantum Hall disk confined by some external electrostatic potential. For simplicity, assume the corresponding edge modes to be described by a 1D field theory for a complex fermionic field $\psi(x,t)$. There would be more than one fermion in the case of several edge channels, and no fundamental fermion altogether for fractional quantum Hall droplets. None of this is crucial for our arguments, since bosonization ultimately relies on the system's global U(1) symmetry, which is present in any case.

The action functional for $\psi$ (in units where $\hbar=1$) is
\be
\label{e1}
S_{\text{f}}[\psi,\psi^{\dagger}]
\equiv
\int\dd t\,\dd x\,
\Big(i\psi^{\dagger}\der_t\psi-\cH_{\text{f}}[\psi,\psi^{\dagger}](x)\Big).
\ee
Here $\cH_{\text{f}}$ is some real local Hamiltonian density that depends on the fermion $\psi$, its conjugate $\psi^{\dagger}$ and their derivatives up to some finite order, all evaluated at the same point $x$, so that $\cH_{\text f}[\psi,\psi^\dagger](x) = \cH_{\text f}\left(\psi(x),\der_x\psi(x),\der^2_x\psi(x)\dots,\psi^\dagger(x),\der_x\psi^\dagger(x),\der^2_x\psi^\dagger(x)\dots\right)$. Setting to zero the variation of the action \eqref{e1}, one finds the fermionic equation of motion
\be
\label{ss4}
i\der_t\psi(x,t)
=
\frac{\delta H_{\text{f}}}{\delta\psi^{\dagger}(x,t)}
\equiv
\sum_{n=0}^{\infty}(-\der_x)^n\left(\frac{\der\cH_{\text{f}}}{\der(\der_x^n\psi^{\dagger})}\right),
\ee
where $H_{\text{f}}[\psi,\psi^{\dagger}]\equiv\oint\dd x\,\cH_{\text{f}}[\psi,\psi^{\dagger}](x)$ is the fermionic Hamiltonian.\footnote{For brevity, we write $\oint\dd x\equiv\int_0^{2\pi}\dd x$ throughout.} This is also meant to hold as a Heisenberg equation of motion in the full quantum theory, in which case the ordering of operators in $H_{\text{f}}$ matters.

In most setups, the leading part of the dynamics \eqref{ss4} is given by the chiral wave equation $\der_t\psi=-\omega\der_x\psi$ for some angular velocity $\omega$, so that the Hamiltonian density in \eqref{e1} is
\be
\label{e2}
\cH_{\text{f}}[\psi,\psi^\dagger]
=
-i\omega\,\psi^{\dagger}\der_x\psi 
+
\text{small corrections}.
\ee
The corrections on the right-hand side typically include dispersive terms $\psi^{\dagger}\der_x^n\psi$ or interactions $(\psi^{\dagger}\psi)^n$ \cite{Agam:2001eu,Abanov:2005nt,Bettelheim:2006un,Wiegmann:2011np,NardinCarusotto,Nardin:2022mto,Nardin_2023,Nardin:2024dyk}. Whenever $n>1$, such corrections are irrelevant in the sense of the renormalization group, while the leading Hamiltonian in \eqref{e2} is exactly marginal---as it should be, since it describes a conformal field theory. It is worth stressing that such small but finite corrections affect the short-distance behaviour of correlations, making it impossible to use bosonization tools based on vertex operators \cite{Senechal:1999us}. This is why our approach will, instead, be semiclassical \cite{Karabali:1991hm,Iso:1992aa,Das:1991qb,Das:1991uta,Sakita:1996ne}. Furthermore, no attempt will be made here to \i{derive} the full Hamiltonian density $\cH_{{\text{f}}}$, say for the edge modes of a quantum Hall sample: our point is precisely to derive conclusions that hold for \i{any} such Hamiltonian, although we will eventually focus on the special case of Korteweg-de Vries dynamics, for the sake of illustration, in section \ref{sesolitons}.

\medskip
\noindent\textbf{Global U(1) symmetry.}
A crucial property of the action \eqref{e1} is its symmetry under global U(1) transformations $\psi\to e^{i\alpha}\psi$, $\psi^{\dagger}\to e^{-i\alpha}\psi^{\dagger}$ where $\alpha$ is any ($x$-independent) real number. This entails the conservation of the Noether charge
\be
Q
\equiv
\oint\frac{\dd x}{2\pi}\,\rho(x),
\label{e4}
\ee
which is the average of the ($2\pi$-periodic) local charge density
\be
\rho(x)
\equiv
\psi^{\dagger}(x)\psi(x).
\label{e5}
\ee
(If there are $\nu>1$ edge channels, the total charge density is a sum of $\nu$ separate densities \eqref{e5}---one for each channel.) Indeed, the equation of motion \eqref{ss4} implies the continuity equation
\be
\der_t\rho+\der_x j
=
0,
\label{bb1b}
\ee
where the local current density can be deduced from \eqref{ss4}, although it is not essential for what follows:
\be
j(x)
\!=\!
\tfrac{i}{2}\!\sum_{n= 1}^\infty\!\sum_{k=1}^n\!\!\binom{n}{k}\!(-\partial_x)^{k-1}\!\!\left(\!\frac{\partial\cH_{\text{f}}}{\partial(\partial_x^n\psi(x))}\partial_x^{n-k}\psi(x)\!\right)
\!+\text{c.c.}
\label{tt1b}
\ee
Note that the latter depends on the Hamiltonian, while the charge density \eqref{e5} does not, and is thus universal. For example, the current for a conformal Hamiltonian $\cH_{\text{f}}=-i\omega\psi^{\dagger}\der_x\psi$ is just $j=\omega\rho$, but the irrelevant corrections in eq.~\eqref{e2} generally make the constitutive relation linking $j$ to $\rho$ much more complicated. A key point is that such a relation exists: by energy conservation, it can be written as
\be
j(x)
\equiv
\nu\frac{\delta H[\rho]}{\delta\rho(x)}
\equiv
\nu\sum_{n=0}^{\infty}(-\der_x)^n\left(\frac{\der\cH}{\der(\der_x^n\rho)}\right),
\label{ttt1b}
\ee
where the normalization factor $\nu>0$ (later identified with the filling fraction) is introduced for future convenience, and
\be
\label{e21}
H[\rho]
\equiv
\oint\dd x\,\cH[\rho](x)
\ee
is some bosonic energy functional whose value is constant along time evolution \eqref{bb1b}. We shall assume that the bosonic Hamiltonian density $\cH[\rho](x)=\cH\big(\rho(x),\der_x\rho(x),\der_x^2\rho(x),...\big)$ is local, in that it only depends on $\rho$ and finitely many of its derivatives, all evaluated at the same point $x$. In the conformal case, the Hamiltonian density would just be $\cH=\omega\rho^2/2\nu$. The latter generally receives subleading (irrelevant) corrections, similarly to its fermionic counterpart \eqref{e2}. Again, we will not attempt to derive the specific Hamiltonian relevant for quantum Hall samples: all that matters is that such a Hamiltonian exists.

\medskip
\noindent\textbf{Classical bosonization.}
Bosonization consists in viewing the continuity equation \eqref{bb1b}, rather than its fermionic parent \eqref{ss4}, as the main equation of motion of the theory. At the classical level, the method rests on the Poincar\'e lemma, just like the many incarnations of particle-vortex duality \cite{Peskin:1977kp,Dasgupta:1981zz,Son:2015xqa,Wang:2015qmt,Metlitski:2015eka}. It boils down to introducing a ($2\pi$-periodic) bosonic field $\phi(x)=\phi(x+2\pi)$ such that\footnote{\label{foowind}Bosonization is commonly phrased in terms of a compact scalar field $\Phi$ that need not be periodic in $x$, which makes it possible to produce the charge $Q$ in \eqref{e10} from a winding configuration $\Phi(x)=Qx+\nu\phi(x)$ such that $\Phi(x+2\pi)-\Phi(x)=2\pi Q$. This perspective is not needed here, as the discussion will focus on a sector with fixed $Q$, \ie fixed winding of $\Phi$.}
\begin{align}
\label{e10}
\rho(x,t)
&\equiv
Q+\nu\,\der_x\phi(x,t),\\
\label{e15b}
j(x,t)
&\equiv
-\nu\,\der_t\phi(x,t),
\end{align}
where the filling fraction $\nu$ is again introduced for later convenience. Then the continuity equation \eqref{bb1b} is automatically satisfied and the charge \eqref{e4} takes the value $Q$, fixed once and for all. Eq.~\eqref{e15b} plays the role of dynamical equation of motion for $\phi$, to be combined with the constitutive relation \eqref{ttt1b} and plugged back in eq.~\eqref{e10} to yield the equation of motion for density. The resulting pair of equations of motion is
\begin{align}
\label{e18a}
\der_t\rho
&=
-\der_xj
=
-\nu\,\der_x\left(\frac{\delta H}{\delta\rho}\right),\\
\label{e18b}
\der_t\phi
&=
-\frac{1}{\nu}j
=
-\frac{\delta H}{\delta\rho},
\end{align}
where everything is evaluated at the same point $(x,t)$, omitted for brevity. In the bosonized theory, these are the fundamental equations of motion of the system, with a status similar to that of the initial fermionic equation of motion \eqref{ss4}. We emphasize that this holds regardless of the Hamiltonian in eq.~\eqref{e1}, well beyond the realm of conformal field theory. 

At fixed charge $Q$, the equations of motion \eqref{e18b} extremize the following bosonic action functional \cite{Floreanini:1987as}:
\be
S[\phi]
\equiv
\int\dd t\,\dd x\Big({-}\frac{\nu}{2}\der_x\phi\,\der_t\phi-\cH[Q+\nu\der_x\phi]\Big).
\label{ss4t}
\ee
The density dynamics \eqref{e18a} then follows from the identification \eqref{e10} between $\rho$ and $\der_x\phi$. We stress that $Q$ here is seen as an external parameter whose value is fixed: it is not meant to be varied in \eqref{ss4t}, and it has no equation of motion. As a consequence, the bosonic action \eqref{ss4t} admits a gauge redundancy $\phi(x,t)\to\phi(x,t)+\alpha(t)$ that was absent in the fermionic theory \eqref{e1}.

For future use, let us derive the Hamiltonian action corresponding to \eqref{ss4t}. This requires finding the momentum conjugate to $\phi$, which is essentially the density \eqref{e5}:
\be
\frac{\delta S}{\delta\partial_t\phi(x)}
=
-\frac{\nu}{2}\partial_x\phi(x)
\stackrel{\text{\eqref{e10}}}{=}
-\frac{1}{2}\big(\rho(x)-Q\big).
\ee
Because the action \eqref{ss4t} is linear in time derivatives, in contrast to nonchiral systems, this momentum is independent of the `velocity' $\der_t\phi$, implying that the relation \eqref{e10} between $\rho(x)$ and $\der_x\phi(x)$ must be enforced by a Lagrange multiplier $u(x,t)$ in the Hamiltonian action \cite{HenneauxTeitelboim}. The latter thus reads
\be
S_{\text{H}}[\phi, \rho, u]
=
\int \dd t\,\dd x
\Big(
{-}\frac{1}{2}\rho\,\partial_t\phi
- \cH[\rho]
+ u\big(Q + \nu\der_x \phi - \rho \big)
\Big),
\label{sh}
\ee
where all three variables $\phi,\rho,u$ are functions of $(x,t)$ and a total time derivative $\int\dd t\,Q\der_t\phi$ was discarded since it does not contribute to the dynamics. The action \eqref{sh} is `Hamiltonian' in that it is a functional of paths in the phase space of pairs of functions $(\phi(x),\rho(x))$ on $S^1$. However, one should keep in mind that the constraint \eqref{e10} enforced by $u$ heavily reduces the dimension of the physical phase space, which is spanned only by functions $\rho(x)$ with a fixed average $Q$. (The constraint \eqref{e10} does not fix the zero-mode of $\phi$, but the latter is pure gauge so it will also eventually drop out of the reduced phase space of the theory \eqref{sh}.) This reduction is represented by the dashed arrows in fig.~\ref{FIFLOW}, and will reverberate throughout our discussion of Berry phases and Lie-Poisson equations. In fact, the Berry phase shown in yellow in fig.~\ref{FIFLOW} will be the symplectic area in the reduced phase space of densities alone.

One aspect of bosonization that is missing in eqs.~\eqref{e10}--\eqref{e15b} is the explicit link between the fermionic field $\psi$ and its bosonic counterpart $\phi$. In short, the latter is the phase of the former. This is embodied by the famous relation $\psi\sim\,:\!e^{i\phi}\!:$ in conformal field theory, where the fermion is a vertex operator---see \eg \cite{Senechal:1999us}. (The colons denote bosonic normal ordering, and all normalizations are omitted.) The full story, however, is more involved, especially when the irrelevant corrections in the Hamiltonian \eqref{e2} are nonzero, which affects the short-distance behaviour of correlators of $\psi$. In that more general case, the link between $\psi$ and $\phi$ stems again from conservation of the U(1) charge \eqref{e4}. Seeing the latter as the $L^2$ squared norm of the field $\psi(x,t)$, time evolution is represented by a time-dependent unitary kernel that implements area-preserving deformations of the underlying 2D quantum liquid \cite{Das:1991qb,Das:1991uta,Karabali:1991hm,Iso:1992aa,Sakita:1996ne,Cappelli1}. The phase of this kernel is, up to normalization, the scalar field $\phi(x,t)$ of eqs.~\eqref{e10}--\eqref{e15b} \cite{Delacretaz:2022ocm}. In what follows, we will therefore view shifts $\phi(x,t)\to\phi(x,t)+\text{const}$ of the bosonic field as U(1) rotations of the fermion $\psi$.

\medskip
\noindent\textbf{U(1) current algebra.}
Note from the action \eqref{sh} that $\phi$ and $\rho$ are canonically conjugate. Indeed, the kinematical term of \eqref{sh} is a field-theoretic version of $\int\dd t\,p\,\der_tq$, with $\rho$ playing the role of `momentum' and $\phi$ that of `position'. It follows that $\phi,\rho$ satisfy the Poisson bracket\footnote{The Poisson bracket $\{\cdot,\cdot\}$ is antisymmetric and has nothing to do with an anticommutator. In the quantum theory, the commutator satisfies the quantization rule $[\hat{\cdot},\hat{\cdot}]=i\widehat{\{\cdot,\cdot\}}$, at least at leading order in $\hbar$.}
\be
\{\phi(x),\rho(y)\}
=
-2\delta(x-y),%
~\text{\ie}~%
\{\phi_m,\rho_n\}
=
-\frac{1}{\pi}\delta_{m+n,0}
\label{e16}
\ee
in terms of Fourier modes such that $\phi(x)=\sum_{n\in\ZZ}\phi_n e^{inx}$ and $\rho(x)=\sum_{n\in\ZZ}\rho_n e^{inx}$. Upon reducing the phase space of pairs $(\phi,\rho)$ to the one where the constraint \eqref{e10} holds, the Poisson bracket \eqref{e16} induces a Dirac bracket such that the nonzero modes of $\rho(x)$ fail to commute. This is similar to noncommuting planar coordinates in the presence of a magnetic field, when the phase space of a charged particle is reduced to the subspace where magnetic momenta vanish---\ie to the classical counterpart of the lowest Landau level \cite{Dunne:1989hv}. In the case at hand, the nonzero modes of the constraint \eqref{e10} can be written as $\chi_n\equiv in\nu\phi_n-\rho_n\approx0$ for all $n\neq0$.\footnote{The standard symbol $\approx$ denotes equalities on the constraint surface \cite{HenneauxTeitelboim}; no approximation is involved.} These are second-class constraints since the matrix of their Poisson brackets \eqref{e16} is nondegenerate,
\be
C_{mn}
\equiv
\{\chi_m,\chi_n\}
=
\frac{2im\nu}{\pi}\delta_{m+n,0}
\qquad
\text{for }m\neq0,\,n\neq0.
\label{s6b}
\ee
Denoting the inverse of this matrix by $C^{mn}=\frac{i\pi}{2m\nu}\delta_{m+n,0}$ for nonzero $m,n$, the Dirac bracket of nonzero modes of density is defined as \cite{HenneauxTeitelboim}
\be
\{\rho_m,\rho_n\}_{\text{D}}
\equiv
\{\rho_m,\rho_n\}
-
\sum_{k,\ell\in\ZZ^*}\{\rho_m,\chi_k\}C^{k\ell}\{\chi_{\ell},\rho_n\}.
\ee
The canonical bracket \eqref{e16}, along with the inverse of the constraint matrix \eqref{s6b}, then yields
\be
\begin{split}
\{\rho_m,\rho_n\}_{\text{D}}
&=
\frac{in\nu}{2\pi}\delta_{m+n,0},\\%
\{\rho(x),\rho(y)\}_{\text{D}}
&=
-\nu\,\der_x\delta(x-y).
\end{split}
\label{e19}
\ee
This is the noncommutative density algebra announced above. It is a U(1) current algebra that plays a key role in what follows, with a central charge equal to the filling fraction $\nu$. As before, the actual value of $\nu$ is unimportant in \eqref{e19}, since it can be set to one by rescaling $\rho$. The advantage of keeping $\nu$ explicit is to ease comparison with section \ref{SAFFI} and with the quantum Hall literature: see \eg \cite{Cappelli1,Cappelli:1993ei} or \cite[sec.~6.1.3]{Tong}.

It is worth comparing such a classical approach to the usual derivation of the U(1) current algebra \eqref{e19} in quantum field theory. Assuming antiperiodic boundary conditions for the fermion $\psi(x)=-\psi(x+2\pi)$, one would view the latter as an operator with a Fourier expansion $\hat\psi(x)=\sum_{p\in\ZZ+1/2}e^{ipx}\hat c{}_p$. Here each $\hat c{}_p$ is a canonically normalized fermionic annihilation operator with momentum $p$, so that $\{\hat c{}_p,\hat c{}_q\}=\delta_{pq}$. Further assuming that the ground state $|0\rangle$ is a filled `Fermi sea' such that
\be
\hat c{}^{\dagger}_p|0\rangle
=0\quad\forall\,p<0,
\qquad
\hat c{}_p|0\rangle
=0\quad\forall\,p>0,
\ee
one would define normal-ordered fermionic bilinears as
\be
:\!\hat c{}^{\dagger}_p\hat c{}_q\!:\,
\equiv
\begin{cases}
\hat c{}^{\dagger}_p\hat c{}_q & \text{if }q>0,\\
-\hat c{}_q\hat c{}^{\dagger}_p & \text{if }q<0.
\end{cases}
\ee
Then the normal-ordered local operator corresponding to the density \eqref{e5} would have commutators that satisfy eq.~\eqref{e19} with $\nu=1$. More generally, the algebra \eqref{e19} holds for the normal-ordered density operator of an integer number $\nu\geq1$ of fermionic fields. While such a quantum derivation is perhaps more familiar to most readers, we emphasize that our approach will instead be classical, which is why eqs.~\eqref{e19} were derived in terms of Dirac brackets rather than commutators.

Finally, a clarification is in order regarding the zero-mode of the constraint \eqref{e10}, namely $\chi_0\equiv Q-\rho_0\approx0$ where the value of $Q$ is fixed once and for all. This is a first-class constraint that generates the gauge transformations $\phi_0\to\phi_0+\alpha(t)$ mentioned below eq.~\eqref{ss4t}, in contrast to the second-class constraints $\chi_{n\neq0}$ introduced above eq.~\eqref{s6b}. It says that the zero-mode $\phi_0$ is a pure gauge degree of freedom, and should be fixed by a gauge condition when taking the action \eqref{ss4t} as a starting point. Thus, the fully reduced phase space of \eqref{ss4t} is the one where all $\phi_n$'s and $\rho_0$ are fixed, while the $\rho_n$'s with $n\neq0$ are free. This is nothing but the space of densities with fixed average mentioned above, and we will see that it coincides with a coadjoint orbit of the affine U(1) group. Remarkably, the phase shift $\Delta\phi$ in fig.~\ref{FIFLOW} will precisely affect the zero-mode $\phi_0$, eventually corresponding to a potentially observable phase rotation of the fermionic field $\psi$. This is similar to the physics of the phase of a wave function: the overall phase is pure gauge, unless one makes the rotated wave function interfere with itself---in which case the phase becomes relative, hence physically observable. In the same vein, we expect the phase shift of $\phi$ to contribute \eg to quantum Hall edge interferometry \cite{Bartolomei:2020qfr,Feldman,Frigerio}, for both $\nu=1$ and $\nu\neq1$, including fractional fillings.

\subsection{Affine U(1) group}
\label{SAFFI}

Here we review the definition of the affine U(1) group and its algebra \eqref{e19}, then introduce Lie-Poisson dynamics on its cotangent bundle. The resulting equations of motion will coincide with \eqref{e18a}--\eqref{e18b}. We refer again to appendix \ref{appa} or \cite[sec.~2 and app.~A]{Oblak:2020jek} for generalities on Lie groups and Lie-Poisson dynamics.

\medskip
\noindent\textbf{Affine U(1) group and current algebra.}
The (universal cover of the) \i{affine U(1) group}, also known as the U(1) Kac-Moody group, is the set $C^{\infty}(S^1)\times\RR$ that consists of pairs $(\phi,a)$, where $\phi(x)$ is a smooth $2\pi$-periodic real function as in eqs.~\eqref{e10}--\eqref{e15b}, and $a$ is a real number.\footnote{As mentioned in footnote \ref{foowind}, the `phase field' $\phi(x)$ is noncompact for simplicity. This is why the relevant group is the \i{universal cover} of the affine U(1) group, rather than the multiply connected affine U(1) group.} It is endowed with the binary operation
\be
\label{e24}
(\phi_1,a_1)\cdot(\phi_2,a_2)
\equiv
\Big(\!\phi_1\!+\!\phi_2,a_1\!+\!a_2\!+\!\oint\!\frac{\dd x}{4\pi}\phi_1(x)\der_x\phi_2(x)\!\Big),
\ee
which makes it a group: the operation \eqref{e24} is associative, the identity element is the pair $(0,0)$, and inversion is given by $(\phi,a)^{-1}=(-\phi,-a)$. Note that this group is nearly Abelian: it only fails to be such due to the central extension $\oint\phi_1\der_x\phi_2$ on the right-hand side. It can be seen as an infinite-dimensional Heisenberg group.

The Lie algebra of the affine U(1) group is the U(1) current algebra, which consists again of pairs\footnote{We abuse notation: strictly speaking, different symbols should denote group elements $(\phi,a)$ and Lie algebra elements, since the latter are by definition tangent vectors of the group at the identity. Here the difference is minor because the group operation \eqref{e24} is just vector addition, up to the central extension. In this sense, Lie algebra elements are `small' group elements, which justifies our notation.\label{foot1}} $(\phi,a)\in C^{\infty}(S^1)\oplus\RR$. It is acted upon by the group according to the adjoint representation
\be
\label{ADD}
\begin{split}
\text{Ad}_{\phi_1}(\phi_2,a_2)
&\equiv
\frac{\der}{\der t}\Big|_0\Big[(\phi_1,0)\cdot(t\phi_2,ta_2)\cdot(\phi_1,0)^{-1}\Big]\\
&=
\Big(\phi_2, a_2+\oint\frac{\dd x}{2\pi}\,\phi_1\der_x\phi_2\Big),
\end{split}
\ee
where we lighten the notation by omitting the central entry in the subscript, since central terms act trivially. The corresponding Lie bracket is the differential of the adjoint group action at the identity, namely
\be
\begin{split}
\Big[(\phi_1,a_1),(\phi_2,a_2)\Big]
&\equiv
\frac{\der}{\der t}\Big|_0
\text{Ad}_{t\phi_1}(\phi_2,a_2)\\
&=
\Big(0,\oint\frac{\dd x}{2\pi}\,\phi_1\der_x\phi_2\Big).
\end{split}
\label{e12}
\ee
It is common to rewrite this in a basis of (complex) Fourier modes: letting $\rho_m\equiv(e^{imx},0)$ and $Z\equiv (0,2\pi)$, the bracket \eqref{e12} yields the commutation relations
\be
\label{e27}
[\rho_m,\rho_n]
=
\frac{in}{2\pi}Z\,\delta_{m+n,0}.
\ee
We already encountered these in the Dirac brackets \eqref{e19}, with a central charge represented by the constant $Z=\nu$.

\medskip
\noindent\textbf{Momentum space and coadjoint representation.}
In Lie-Poisson systems, `momentum space' is actually the dual vector space of a Lie algebra. In the case at hand, this dual consists of pairs $(\rho,\nu)$, where $\rho(x)$ is again a (smooth) $2\pi$-periodic function and $\nu$ is a real number. The former will eventually be interpreted as the (classical) density \eqref{e5}, while the latter is a central charge as in the Dirac brackets \eqref{e19}, so it will eventually be interpreted as the (possibly fractional) number of edge channels. To see in what sense each such pair is in the `dual' of the Lie algebra, use it to define a linear form
\be
\label{e14}
\big<(\rho,\nu),(\phi,a)\big>
\equiv
\oint\frac{\dd x}{2\pi}\,\rho(x)\phi(x)+\nu a
\ee
for any $(\phi,a)$ in the current algebra. Now, any Lie group acts on the dual of its algebra by the \i{coadjoint representation}, \ie the dual of the adjoint. Here the adjoint is given by \eqref{ADD}, so the coadjoint action reads
\be
\label{COAD}
\text{Ad}^*_{\phi}(\rho,\nu)
\equiv
\underbrace{\big<(\rho,\nu),\text{Ad}_{\phi^{-1}}(\cdot)\big>}_{\text{`dual of adjoint'}}
=
\big(\rho+\nu\der_x\phi,\nu\big)
\ee
where we implicitly used the pairing \eqref{e14}. One may view this action as a gauge transformation of the `gauge field' $\rho(x)\dd x$ with a gauge parameter $\phi(x)$. The corresponding coadjoint representation of the Lie algebra is obtained by differentiation, similarly to the bracket \eqref{e12}:
\be
\label{e17}
\text{ad}^*_{\phi}(\rho,\nu)
\equiv
\frac{\der}{\der t}\Big|_0
\text{Ad}^*_{t\phi_1}(\rho,\nu)
=
(\nu\der_x\phi,0),
\ee
which is akin to an infinitesimal gauge transformation of $\rho$.

Importantly, eqs.~\eqref{COAD}--\eqref{e17} exhibit that the central charge $\nu$ is unaffected by the coadjoint action; once chosen, its value is fixed once and for all. The same is true of the average value \eqref{e4} of $\rho(x)$, while the nonzero Fourier modes of $\rho(x)$ can be shifted at will owing to eq.~\eqref{COAD}. As a consequence, any coadjoint orbit of the affine U(1) group,
\be
\cO_{Q,\nu}
\equiv
\Big\{%
\Ad^*_{\phi}(Q,\nu)=(Q+\nu\,\der_x\phi,\nu)
\,\Big|\,
\phi\in C^{\infty}(S^1)
\Big\},
\label{s9b}
\ee
is uniquely labelled by the average $Q$ of $\rho(x)$ and the central charge $\nu$, here assumed to be nonzero. (For $\nu=0$, each coadjoint orbit consists of a single point since the coadjoint action \eqref{COAD} is trivial.) This orbit is the one announced at the very end of section \ref{SEBO}: it is the fully reduced phase space of the bosonized theory \eqref{sh}, consisting of all densities with a given, fixed average $Q$. From this perspective, the bosonized action \eqref{ss4t} is a `geometric action' for the affine U(1) group \cite{Alekseev:1988ce,Alekseev:1990mp,Barnich:2017jgw,Cotler:2018zff}.

\begin{figure}[t]
\includegraphics[width=.23\textwidth]{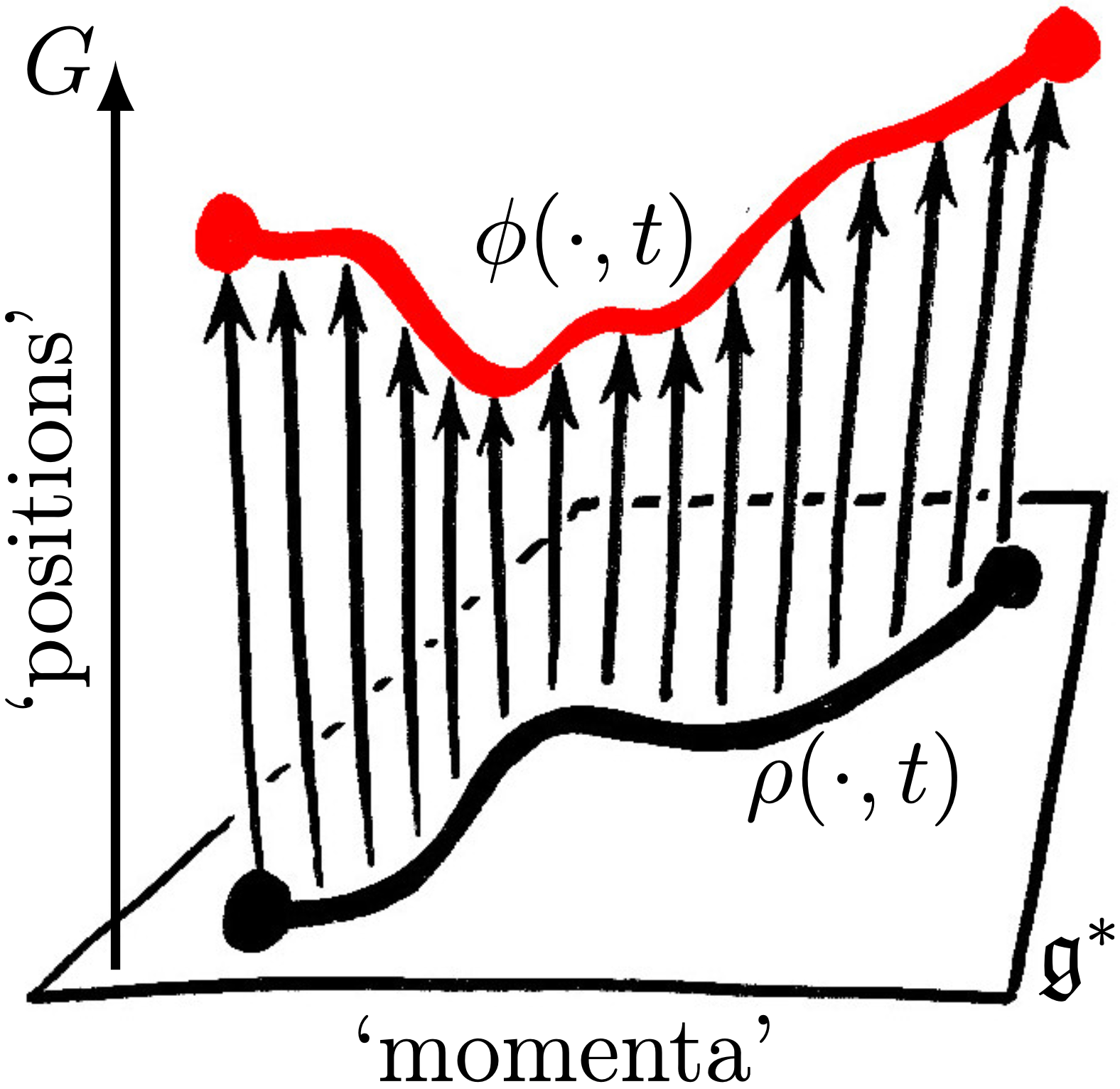}
\caption{The phase space $G\times\mg^*$ of any Lie-Poisson equation involves `position space' $G$ and `momentum space' $\mg^*$. In the case of the bosonized equations of motion \eqref{e18a}--\eqref{e18b}, $G$ is an affine U(1) group. The Lie-Poisson equation \eqref{ss85} and its reconstruction \eqref{ss75} then determine paths in $G\times\mg^*$ that can be built in two steps: (i) find $\rho(x,t)$ by integrating the momentum equation \eqref{ss85} (black curve); (ii) obtain $\phi(x,t)$ from the relation \eqref{ss75} between velocity and momentum (red curve). Identical arguments apply to any system of Lie-Poisson equations \eqref{e28a}--\eqref{e28b}.}
\label{fix}
\end{figure}

\medskip
\noindent\textbf{Lie-Poisson dynamics for the affine U(1) group.}
Lie-Poisson equations of motion can be defined for any Lie group $G$ with a Lie algebra $\mg$: one views the product $G\times\mg^*$ as a phase space, identifying elements $g\in G$ with `configurations' and elements $p\in\mg^*$ with `momenta'; see fig.~\ref{fix}. These are canonically conjugate, leading to the aforementioned Poisson bracket \eqref{e16} in the affine U(1) case. Assuming the Hamiltonian $H=H(p)$ only depends on momenta, the resulting equations of motion turn out to involve the coadjoint representation \eqref{e17}:
\begin{align}
\label{e28a}
\der_tp(t)
&=
-\ad^*_{\der H/\der p}(p),\\
\label{e28b}
\der_tg(t)\,g(t)^{-1}
&=
-\frac{\der H}{\der p},
\end{align}
where $\der_tg\,g^{-1}$ is the (right) logarithmic derivative of the path $g(t)$ in $G$; see the discussion around eq.~\eqref{e79} in appendix \ref{appa2} for details.

Let us unpack these formulas for the affine U(1) group. In that case, `momenta' are densities $\rho(x)$ on a circle, so let \eqref{e21} be a local Hamiltonian functional as in the constitutive relation \eqref{ttt1b}. Since \eqref{e21} is a real function on the dual of the current algebra, its differential $\delta H/\delta\rho(x)$ may be viewed as an algebra element. The latter may therefore act on dual vectors according to the coadjoint representation \eqref{e17}, as in $\ad^*_{\delta H/\delta\rho}(\rho,\nu)$. This is precisely the time derivative of $p=(\rho,\nu)$ in the Lie-Poisson equation \eqref{e28a}, which thus becomes
\be
(\der_t\rho,\der_t\nu)
=
-\ad^*_{\frac{\delta H}{\delta\rho}}(\rho,\nu)
\stackrel{\text{\eqref{e17}}}{=}
\bigg(-\nu\,\der_x\Big(\frac{\delta H}{\delta\rho}\Big),0\bigg).
\label{ss85}
\ee
Here the condition $\der_t\nu=0$ confirms that the central charge is constant, while the equation of motion for $\rho$ reproduces the continuity equation \eqref{bb1b} in its form \eqref{e18a}. It remains to recover eq.~\eqref{e18b} for $\phi$ from eq.~\eqref{e28b}.

To this end, suppose one has solved the momentum equation of motion \eqref{ss85}. In the language of the phase space $G\times\mg^*$, this means one now has a path in momentum space $\mg^*$. The issue is to find the corresponding motion in configuration space $G$, \ie a time-dependent pair $(\phi(x,t),a(t))$, tracing a path in the affine U(1) group, that satisfies eq.~\eqref{e28b}. We shall refer to this step as the \i{reconstruction} of a Lie-Poisson equation: it amounts to reconstructing the time-dependent position of a particle from the knowledge of its momentum, as in fig.~\ref{fix}. As usual in Hamiltonian mechanics, reconstruction is provided by the velocity $\delta H/\delta\rho$, where `velocity' should now be understood as the (right) logarithmic derivative determined by the group operation \eqref{e24}:
\be
\begin{split}
&\frac{\der}{\der\tau}\Big|_{\tau=t}\Big[\big(\phi(x,\tau),a(\tau)\big)\cdot\big(\phi(x,t),a(t)\big)^{-1}\Big]\\
&\stackrel{\text{\eqref{e24}}}{=}
\Big(\der_t\phi,\der_ta-\oint\frac{\dd x}{4\pi}\der_x\phi\der_t\phi\Big)
\stackrel{\text{\eqref{e28b}}}{=}
\Big(-\frac{\delta H}{\delta\rho(x)},0\Big).
\end{split}
\label{ss75}
\ee
Here the equation of motion for $\phi$ reproduces the desired relation \eqref{e18b}, completing the coincidence between bosonization and Lie-Poisson dynamics on the affine U(1) group. But an extra piece of information in \eqref{ss75} is the equation of motion for the central term $a(t)$, which is invisible in the field-theoretic approach of section \ref{SEBO}. As we shall see, the dynamics of $a(t)$ in \eqref{ss75} contributes to the shift of $\phi$ when $\rho(x,t)$ is periodic in time.

This concludes the derivation of Lie-Poisson dynamics for the affine U(1) group, and the coincidence with section \ref{SEBO} is manifest: in addition to the aforementioned momentum equation \eqref{ss85} that coincides with the density dynamics \eqref{e18a}, the reconstruction equation \eqref{ss75} coincides with the phase dynamics \eqref{e18b}. In this sense, bosonization says that the phase field $\phi$ is an element of an affine U(1) group, and that the density \eqref{e5} is an element of the dual of the U(1) current algebra. The link \eqref{e10} between $\rho$ and $\der_x\phi$ is finally found by noting that the solution of the Lie-Poisson equation \eqref{ss85} is entirely contained in a coadjoint orbit \eqref{s9b}: $\big(\rho(x,t),\nu\big)=\text{Ad}^*_{\phi}(Q,\nu)=\big(Q+\nu\,\der_x\phi(x,t),\nu\big)$.

\section{Berry phases from periodic waves}
\label{SEBERR}

In this section, we prove that periodic density waves, such that $\rho(x,t+T)=\rho(x,t)$ for some period $T$, give rise to Berry phases through the field $\phi$. Such phases can be written in closed form in terms of the wave profile $\rho(x,t)$ and the Hamiltonian \eqref{e21} that governs its dynamics, and they are expected to be observable \eg in quantum Hall edge interferometers \cite{Bartolomei:2020qfr,Feldman,Frigerio}. In addition, the Berry phases act as a diagnostic of edge mode dispersion and/or nonlinearity, in that they reduce to the `trivial' average of the wave profile in the conformal limit.

The argument parallels early works on geometric phases in symplectic geometry \cite{Marsden90,Montgomery}, as well as a derivation first presented in \cite{Oblak:2020jek} that was similarly concerned with periodic waves. 
In fact, the approach applies to any Lie group (finite- or infinite-dimensional, with or without central extensions), so it is reviewed in detail in appendix \ref{appb}. The present section will focus 
on the application of the method to the specific case of periodic waves in the affine U(1) group. The plan is to first explain in what sense coherent states of the affine U(1) group can pick up Berry phases, then show how such phases appear in the shift \eqref{s125} of $\phi$ after one period of $\rho$, and finally turn to various illustrative examples that include periodic Korteweg-de Vries solitons. Despite the resulting similarity to \cite{Oblak:2020jek}, we stress that the underlying group, the computation of the phases and their physical effects are all different from those encountered in that earlier reference.

\subsection{Berry phases of coherent states\\in the affine U(1) group}

Let us show how periodic density perturbations give rise to Berry phases. For a similar and more detailed derivation in conformal field theory, see \cite{Oblak:2017ect}; for generalities on adiabatic drives and Berry phases for coherent states, see \eg \cite{jordan1988berry,Bose,Boya,Zalesny:2000zz}.

\medskip
\noindent\textbf{Coherent states of density.}
Consider a Hilbert space where the U(1) current algebra \eqref{e27} is represented by operators $\hat\rho_m=\hat\rho^{\dagger}_{-m}$, with a central charge operator $\hat Z=i\nu\hat I$. This is a unitary representation that can be decomposed as a direct sum of irreducibles, each of which admits a unique highest-weight state $|Q\rangle$ such that
\be
\hat\rho_0|Q\rangle=Q|Q\rangle,
\qquad
\hat\rho_m|Q\rangle=0\quad\text{if }m>0.
\label{t105}
\ee
Here one should think of the $\hat\rho_m$'s as modes of the Hermitian density operator $\hat\rho(x)=\sum_{n\in\ZZ}\hat\rho_n e^{inx}$ corresponding to the quantization of the classical density \eqref{e5}. Any unitary operator in the affine U(1) group reads
\be
\hat U[\phi]\equiv\exp\Big(i\oint\frac{\dd x}{\sqrt{2\pi}}\,\phi(x)\hat\rho(x)\Big)
\ee
for some classical, real phase field $\phi(x)$. In accordance with the commutators \eqref{e27}, the composition law of such unitaries reproduces the group operation \eqref{e24}:
\be
\hat U[\phi_1]\circ \hat U[\phi_2]
=
e^{i\nu\oint\frac{\dd x}{4\pi}\phi_1\der_x\phi_2}\,\hat U[\phi_1+\phi_2],
\label{s105}
\ee
which will be important below. Each state of the form $\hat U[\phi]|Q\rangle$ is a coherent state, obtained by acting on the highest-weight state \eqref{t105} with a finite affine U(1) transformation. From the perspective of geometric quantization, the wave function of $\hat U[\phi]|Q\rangle$ lives on the coadjoint orbit \eqref{s9b} and is localized at the point whose classical density is $\rho(x)=Q+\nu\der_x\phi(x)$.

\medskip
\noindent\textbf{Berry phases.}
Now add some time-dependent dynamics: choose an unperturbed Hamiltonian $\hat H_0$ that admits $|Q\rangle$ as an eigenstate (with zero energy without loss of generality), and that commutes with $\hat\rho_0$ but not with the $\hat\rho_n$'s with $n\neq0$. Then choose a sequence $\phi(x,t)$ of phase field configurations, and impose that the system's Hamiltonian at time $t$ is $\hat H(t)=\hat U[\phi(\cdot,t)]\hat H_0\hat U[\phi(\cdot,t)]^{-1}$. Assuming the perturbation due to $\phi(x,t)$ is slow enough \cite{Born1928,Avron:1998th}, any initial coherent state $|\psi(0)\rangle=\hat U[\phi(\cdot,0)]|Q\rangle$ evolves at time $t$ into a coherent state $|\psi(t)\rangle\sim e^{i\alpha(t)}\hat U[\phi(\cdot,t)]|Q\rangle$ with some overall phase $\alpha(t)$. The latter is a pure gauge parameter, unless the Hamiltonian is periodic, \ie $\hat H(T)=\hat H(0)$. In that case, the phase in $|\psi(T)\rangle=e^{i\alpha(T)}|\psi(0)\rangle$ is a measurable Berry phase \cite{Berry}. Its expression is
\be
\begin{split}
\alpha(T)
&=
i\int_0^T\dd t\,\langle Q|\hat U[\phi(\cdot,t)]^{-1}\der_t\hat U[\phi(\cdot,t)]|Q\rangle\\
&=
-\nu\int\frac{\dd t\,\dd x}{4\pi}\der_x\phi\,\der_t\phi,
\end{split}
\label{s11b}
\ee
where we used the composition law \eqref{s105} and the properties \eqref{t105} of $|Q\rangle$, and assumed without loss of generality that the zero-mode $\phi_0(0)=\phi_0(T)$ is periodic since everything commutes with $\hat\rho_0$. Note that the Berry phase thus coincides with the first term of the bosonized action \eqref{ss4t}, confirming that the phase is actually a symplectic flux on a coadjoint orbit \eqref{s9b}. For the same reason, the Berry phase \eqref{s11b} is independent of the time-dependent zero-mode of $\phi(x,t)$, as long as $\phi(x,t+T)=\phi(x,t)$: the phase only depends on the derivative $\der_x\phi(x,t)$, \ie on the time-dependent classical density $\rho(x,t)=Q+\nu\der_x\phi(x,t)$.

It is worth noting that Berry phases such as \eqref{s11b} can, in principle, be produced by subjecting quantum Hall droplets to suitable families of adiabatic deformations. Namely, in polar coordinates with an angle $x$, consider the area-preserving deformation
\be
(r^2,x)\mapsto(r^2+\ell^2\der_x\phi(x),x)
\label{s115b}
\ee
of a quantum Hall disk with magnetic length $\ell$. (This is, incidentally, the kind of deformation that was used to produce fig.~\ref{FIWAVE}.) The operators implementing such transformations on the edge field theory turn out to satisfy a U(1) current algebra \eqref{e19} whose central charge is the Hall conductance \cite{Cappelli1,Cappelli:1993ei}.  One could thus, theoretically, act on a quantum Hall ground state with time-dependent deformations \eqref{s115b}, whereupon the many-body wave function would pick up a Berry phase \eqref{s11b}. In practice, it seems unlikely that one will ever control decoherence effects enough to measure that Berry phase \eqref{s11b} directly. This is the advantage of the semiclassical approach of this work: the phase \eqref{s11b} will appear in a classical dynamical system, without external drive, requiring neither adiabaticity nor coherence.

\subsection{Phase shifts due to chiral density waves}

As announced, the presence of Berry phases \eqref{s11b} in the dynamics \eqref{e18a}--\eqref{e18b} rests on the Lie-Poisson structure of these equations of motion. The argument is presented in detail (and for any Lie group) in appendix \ref{appb}. It can be summarized as follows.

\medskip
\noindent\textbf{Shift of $\boldsymbol\phi$ as a sum of phases.}
Suppose $\rho(x,t)$ is a solution of \eqref{e18a}--\eqref{e18b} that is periodic in time, so that $\rho(x,t+T)=\rho(x,t)$ for some period $T$. Then, it is certainly the case that the U(1) field $\phi$ in  eq.~\eqref{e18b} satisfies
\be
\label{s135}
\Delta\phi
\equiv
\phi(x,T)-\phi(x,0)
=
-\oint_0^T\dd t\frac{\delta H}{\delta\rho(x,t)}.
\ee
The phase shift $\Delta\phi$ defined in this way is actually independent of $x$: this is readily verified by taking the derivative of \eqref{s135} to find
\be
\begin{split}
\der_x\Delta\phi
&\stackrel{\text{\color{white}{(12)}}}{=}
-\oint_0^T\dd t\,\der_x\left(\frac{\delta H}{\delta\rho}\right)\\
&\stackrel{\text{\eqref{e18a}}}{=}
\oint_0^T\dd t\,\der_t\rho
=
\rho(x,T)-\rho(x,0)
=
0.
\end{split}
\label{t135}
\ee
Thus, it is certainly true that the U(1) field picks up an overall phase $\Delta\phi$ at the end of one period of the density profile $\rho(x,t)$.

What is less obvious is that the phase shift \eqref{s135} can be seen as the sum of a dynamical phase and a Berry phase---at least when the charge \eqref{e4} does not vanish. To understand why this is so, rewrite eq.~\eqref{s135} with nonzero charge $Q$ as
\be
\label{s125}
Q\Delta\phi
=
-\oint\frac{\dd x\,\dd t}{2\pi}\rho\frac{\delta H}{\delta\rho}
+\oint\frac{\dd x\,\dd t}{2\pi}(\rho-Q)\frac{\delta H}{\delta\rho},
\ee
where the first term on the right-hand side is a classical dynamical phase, associated with an arbitrary, generally inhomogeneous Hamiltonian functional. As for the second term on the right-hand side of \eqref{s125}, use eqs.~\eqref{e10} and \eqref{e18b} to rewrite it in terms of the phase field $\phi$. Thereby recognize the Berry phase \eqref{s11b} and finally write
\begin{empheq}[box=\othermathbox]{equation}
\label{KEY}
\Bigg.
Q\,\Delta\phi
=
-\underbrace{\int\frac{\dd t\,\dd x}{2\pi}\rho\,\frac{\delta H}{\delta\rho}}_{\ds\text{Dynamical}}
-\underbrace{\nu\int\frac{\dd t\,\dd x}{2\pi}\,\der_x\phi\,\der_t\phi.}_{\ds\text{Berry}}
\end{empheq}
We stress once more that the Berry phase here need not involve the dynamical phase field that solves the equation of motion \eqref{e18b}. Instead, all one requires is that $\phi(x,t)$ on the right-hand side of \eqref{KEY} be any function, periodic in both space and time, such that $\rho(x,t)=Q+\nu\der_x\phi(x,t)$. Note that the Berry phase in \eqref{KEY} differs from that in \eqref{s11b} by a factor two. This is not a typo: it has to do with the central entry $a(t)$ in \eqref{ss75}, to which we return below.

\medskip
\noindent\textbf{Berry phases as symplectic fluxes.}
While the splitting \eqref{KEY} clearly exhibits the Berry phase \eqref{s11b}, it also has a geometric interpretation that is deeply tied to the reduction from the phase space of pairs $(\phi,\rho)$ to the coadjoint orbit \eqref{s9b} of densities with a prescribed average. The argument is, again, given in appendix \ref{appb}, but it can be outlined as follows.

\begin{figure}[t]
\includegraphics[width=.25\textwidth]{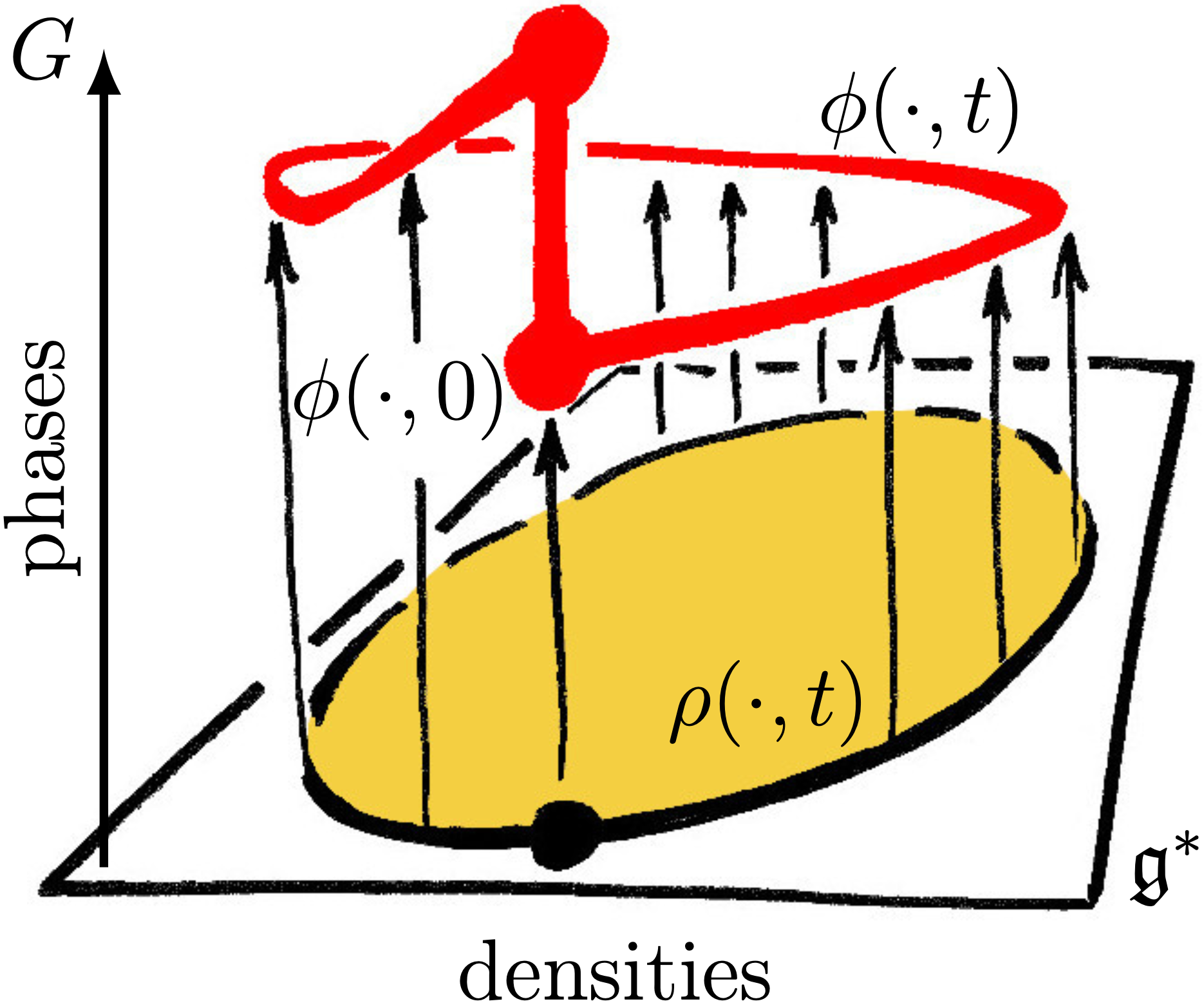}
\caption{As in fig.~\ref{fix}, a solution of the Lie-Poisson equation \eqref{ss85} for the density $\rho(x,t)$ determines a phase field $\phi(x,t)$ through reconstruction \eqref{ss75}. Even if $\rho(x,t)$ is periodic in time, the phase field $\phi(x,t)$ generally is not. The mismatch \eqref{s135} between $\phi(\cdot,T)$ and $\phi(\cdot,0)$ is the sum of a dynamical phase and a Berry phase, the latter being a `symplectic area' (in yellow) on the coadjoint orbit \eqref{s9b}, as in fig.~\ref{FIFLOW}. This turns out to be a general feature of Lie-Poisson equations, in which case `densities' and `phase fields' respectively correspond to `momenta' and `configurations' in the cotangent bundle of a Lie group (see appendices \ref{appa}--\ref{appb}).}
\label{fi4}
\end{figure}

Recall from section \ref{SAFFI} that a Lie-Poisson system consists of a pair of evolution equations: one for a time-dependent `momentum', the other for a time-dependent group element. In the affine U(1) group, these equations are respectively \eqref{ss85} for the density $\rho$ and \eqref{ss75} for the phase $\phi$. Now let the density solving \eqref{ss85} be periodic in time, which occurs for instance (but not only) when $\rho(x,t)=\rho(x-vt)$ is a periodic soliton. Then the phase $\phi$ is fully determined by the `reconstruction equation' \eqref{ss75}, but need not be periodic---as stated indeed by eq.~\eqref{s135}. The red closed curve in fig.~\ref{fi4} thus represents the path
\be
\tilde\phi(x,t)
\equiv
\begin{cases}
\phi(x,t) & t\in[0,T],\\
(\tfrac{t}{T}-1)\phi(x,0)+(2-\frac{t}{T})\phi(x,T) & t\in[T,2T],
\end{cases}
\label{s115}
\ee
whose second half traces the gap between $\phi(\cdot,0)$ and $\phi(\cdot,T)$ while leaving fixed the corresponding density $\rho=Q+\nu\der_x\phi$, since $\der_x\phi(x,0)=\der_x\phi(x,T)$.

This geometric picture provides a symplectic interpretation of the phase shift \eqref{s135}. Indeed, pairs consisting of a phase field $\phi$ and a density $\rho$ may be seen as elements of the cotangent bundle of the affine U(1) group. This bundle is a symplectic manifold with a globally exact symplectic form $\Omega=-\dd\cA$, where the Liouville one-form $\cA$ is effectively the Berry connection in eq.~\eqref{s11b}. Integrating the connection along the closed curve \eqref{s115} yields a Berry phase that can be split in two pieces. Schematically,
\be
\oint\cA
=
\int\limits_{\!\!\!\!\!\!\!\!\!\!\!\!\!\!\!\!\!\!\!\raisebox{-1.4em}{{\tikz{\includegraphics[width=.07\textwidth]{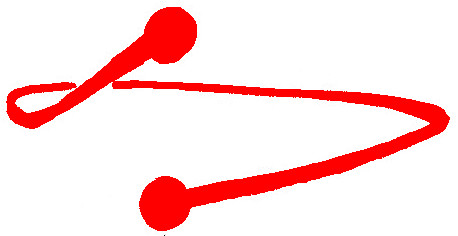};}}}}\cA
\,\,+\,\int\limits_{\!\!\!\!\raisebox{-1.4em}{{\tikz{\includegraphics[width=.01\textwidth]{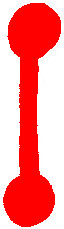};}}}}\cA
\label{t115}
\ee
where the two halves of the path are as in fig.~\ref{fi4}. The first half yields a dynamical phase, the second yields the product $Q\Delta\phi$. Rearranging the equation turns out to reproduce eq.~\eqref{KEY}, with the anticipated interpretation of the two terms on the right-hand side.

Two comments are in order. First, we have been somewhat cavalier regarding the time-dependent central piece $a(t)$ in the reconstruction equation \eqref{ss75}. The latter has not appeared explicitly since we started discussing the phase shift \eqref{s135}. Because its time evolution \eqref{ss75} is nontrivial, it actually does contribute to the shift of $\phi$. It was indeed shown in \cite{Oblak:2020jek} that such central terms give rise to `anomalous phases' on top of the dynamical and Berry phases mentioned around eq.~\eqref{t115}. So why did we seemingly just forget about that effect here? The reason has to do with a remarkable simplification, valid in the affine U(1) group but generally untrue in other centrally extended groups: the anomalous phase due to $a(t)$ in \eqref{ss75} equals the Berry phase \eqref{s11b}, so the latter appears \i{twice} in eq.~\eqref{KEY}---hence the aforementioned factor two mismatch. The end result is that the phase shift \eqref{KEY} seemingly only involves a dynamical phase and a Berry phase, albeit up to crucial normalizations. For completeness, the details of this cancellation are spelled out at the end of appendix \ref{appb}.

Second, note that we only refer to the terms in eq.~\eqref{KEY} as `Berry phase' and `dynamical phase' when the total charge \eqref{e4} is nonzero. This is because the second term on the right-hand side of \eqref{t115} is nonzero only when $Q\neq0$; only then can the phase shift $\Delta\phi$ be thought of in symplectic terms. In this sense, the interpretation is limited to periodic waves whose average density \eqref{e4} does not vanish. There is of course nothing wrong with waves whose average does vanish: they still give rise to a definite phase shift \eqref{s135}, which is still constant by virtue of \eqref{t135}. It just so happens that the phase shift when $Q=0$ apparently admits no Berry phase interpretation.

\subsection{Travelling waves and solitons}
\label{sesolitons}

Here we apply our main result \eqref{KEY} to several special cases. We first consider travelling waves that propagate without change of shape, then turn to quadratic (dispersive) Hamiltonians. Finally, we write eq.~\eqref{KEY} for the Korteweg-de Vries equation that has recently been put forward as a model of nonlinear quantum Hall edge modes \cite{NardinCarusotto,Nardin:2022mto}, and apply it to the special case of cnoidal waves, \ie periodic solitons.

\medskip
\noindent\textbf{Travelling waves.}
A travelling wave is a solution of \eqref{e18a} such that
\be
\rho(x,t)=\rho(x-vt)
\label{TRAVV}
\ee
for some angular velocity $v$. Since $x\sim x+2\pi$ is a coordinate on a circle and $\rho(x)$ is $2\pi$-periodic, any such travelling wave is also periodic in time, with period $T=2\pi/|v|$. It is then a simple matter to apply eq.~\eqref{KEY}: the dynamical phase simplifies since the time integral can now be carried out explicitly, and the Berry phase can be recast entirely in terms of the time-independent profile $\rho(x)$. The end result is
\be
\label{KEYT}
Q\,\Delta\phi
=
-\underbrace{\frac{1}{|v|}\!\oint\dd x\rho(x)\frac{\delta H}{\delta\rho(x)}}_{\ds\text{Dynamical}}
+
\underbrace{\frac{1}{\nu}\frac{v}{|v|}\!\oint\dd x\big(\rho(x)^2-Q^2\big)}_{\ds\text{Berry}}.
\ee
Note how dynamical and Berry phases differ in their scaling with the velocity $v$ \cite{Oblak:2020jek}. Furthermore, the Berry phase now becomes a sort of `variance' of the wave profile on the circle.

Similarly to eq.~\eqref{s135}, the computation of $\Delta\phi$ for travelling waves \eqref{TRAVV} can be achieved without any reference to Berry phases, since eq.~\eqref{s135} yields the `drift velocity'
\be
\label{DRIFT}
\frac{\Delta\phi}{T}
=
-\oint\frac{\dd x}{2\pi}\frac{\delta H}{\delta\rho(x)}.
\ee
This coincides (as it must) with the result \eqref{KEYT} when the latter is divided by the period $T=2\pi/|v|$. Indeed, the equation of motion \eqref{e18a} for travelling waves reduces to the ordinary differential equation $-v\der_x\rho=-\nu\der_x(\delta H/\delta\rho)$, which is trivially integrated to $-v\rho=C-\nu\,\delta H/\delta\rho$ for some integration constant $C$. Any derivative $\delta H/\delta\rho$ may thus be replaced by $(C+v\rho)/\nu$. Performing this replacement in both \eqref{DRIFT} and the earlier result \eqref{KEYT} readily shows that the two formulas agree.

\medskip
\noindent\textbf{Quadratic Hamiltonians.}
Suppose now that the bosonic Hamiltonian \eqref{e21} is quadratic, so that $H[\rho]=\oint\dd x\,\rho(x)D\rho(x)$ for some self-adjoint differential operator $D$ on $L^2(S^1)$. Then the entire Hamiltonian is homogeneous of degree two in $\rho$. The dynamical phase in \eqref{KEY} reduces to a multiple of the conserved energy $E$ of the  wave profile, and eq.~\eqref{KEY} can be recast as
\be
Q\,\Delta\phi
=
-2ET-\nu\int\frac{\dd t\,\dd x}{2\pi}\,\der_x\phi\,\der_t\phi,
\ee
where the factor 2 in $2ET$ stems from the degree of the Hamiltonian. More generally, any Hamiltonian \eqref{e21} that is homogeneous of degree $d$ in $\rho$ leads to a dynamical phase $d\times ET$. In the special case where the Hamiltonian \eqref{e21} has no dispersive part, so $ H[\rho]=\tfrac{\omega}{2\nu}\oint\dd x\,\rho(x)^2$ for some angular velocity $\omega>0$, the equation of motion \eqref{e18a} is solved by $\rho(x,t)=\rho(x-\omega t)$ and eq.~\eqref{s135} for the phase yields
\be
\label{KEYSIMP}
\Delta\phi
=
-2\pi Q/\nu.
\ee
Note how the entire dynamical phase disappears in \eqref{KEYSIMP}, with the only nonzero contribution coming from the $Q^2$ term in the Berry phase \eqref{KEYT}. In this respect, the general phase shift \eqref{KEY} is a diagnostic of dispersion and nonlinearity.

\medskip
\noindent\textbf{Korteweg-de Vries dynamics.}
The Korteweg-de Vries equation is a well known model of shallow-water dynamics \cite{Ockendon}, but it has recently also been rediscovered in the context of nonlinear edge modes of quantum Hall droplets \cite{NardinCarusotto,Nardin:2022mto,Nardin_2023,Nardin:2024dyk}. Its Hamiltonian \eqref{e21} is dispersive and nonlinear:
\be
H[\rho]
\equiv
\oint\dd x\Big(
\frac{\omega}{2}\rho^2
+\frac{\beta}{2}(\der_x\rho)^2
-\frac{\alpha}{6}\rho^3\Big),
\ee
where $\omega,\alpha,\beta$ are constants, $\alpha,\beta$ being nonzero.\footnote{In many references, some or all of these constants are set to one by suitable choices of normalization and units; here we keep the constants explicit to simplify comparison with the literature.} The corresponding density dynamics \eqref{e18a} is the Korteweg-de Vries equation
\be
\label{KDV}
\der_t\rho
=
-\nu\omega\der_x\rho+\nu \beta\der_x^3\rho+\nu \alpha\rho\der_x\rho,
\ee
whose simplest periodic solutions are solitons,\footnote{A minor clarification: the standard Korteweg-de Vries solitons live on the real line rather than the circle, and are definitely periodic neither in space, nor in time. Our terminology here is instead that any travelling wave \eqref{TRAVV} solving a nonlinear wave equation is, by definition, a soliton, even if it lives on a circle. As argued above, any such space-periodic soliton is necessarily also time-periodic.} namely cnoidal waves: then $\rho(x,t)=\rho(x-vt)$ is a travelling wave \eqref{TRAVV} with arbitrary velocity $v$, whose $2\pi$-periodic profile is given by
\be
\label{CNO}
\begin{split}
\rho(x)
&=
\underbrace{\frac{4\beta}{\alpha}\frac{K(m)^2}{\pi^2}(m+1)-\frac{v-\nu\omega}{\nu \alpha}}_{\ds\equiv\lambda(m,v)}\\
&~~~~~\underbrace{-\,\frac{12\beta}{\alpha}m\frac{K(m)^2}{\pi^2}}_{\ds\equiv\mu(m)}\text{sn}^2\Big(\frac{K(m)}{\pi}x\Big|m\Big).
\end{split}
\ee
Here $m\in[0,1[$ is a peakedness parameter, $K(m)$ is the complete elliptic integral of the first kind, $\text{sn}(u|m)$ is the Jacobi elliptic sine function with modulus $m$, and the notations $\lambda,\mu$ are introduced for later convenience. Note that this profile is entirely specified by the soliton velocity $v$ and the peakedness parameter $m$, assuming the couplings $\omega,\alpha,\beta$ are fixed once and for all.

Since the Korteweg-de Vries equation \eqref{KDV} is integrable, each of its solutions may be seen as a superposition of solitons \eqref{CNO} \cite{Novikov}. It is clear that such multi-soliton solutions can be built in a way that ensures periodicity in time: think \eg of two solitons whose dephasing angle is a rational multiple of $2\pi$ \cite{boyd1982theta,boyd1984double}. Then, for any time-periodic solution of \eqref{KDV}, the phase equation \eqref{KEY} applies with the following integrand for the dynamical phase:
\be
\label{DYNI}
\rho\frac{\delta H}{\delta\rho}
=
\omega\rho^2+\beta(\der_x\rho)^2-\frac{\alpha}{2}\rho^3.
\ee
(In writing this we integrated by parts to simplify the right-hand side.) The Berry phase \eqref{s11b} can similarly be evaluated once a periodic wave profile $\rho(x,t)$ is given.

\begin{figure}[t]
\includegraphics[width=.3\textwidth]{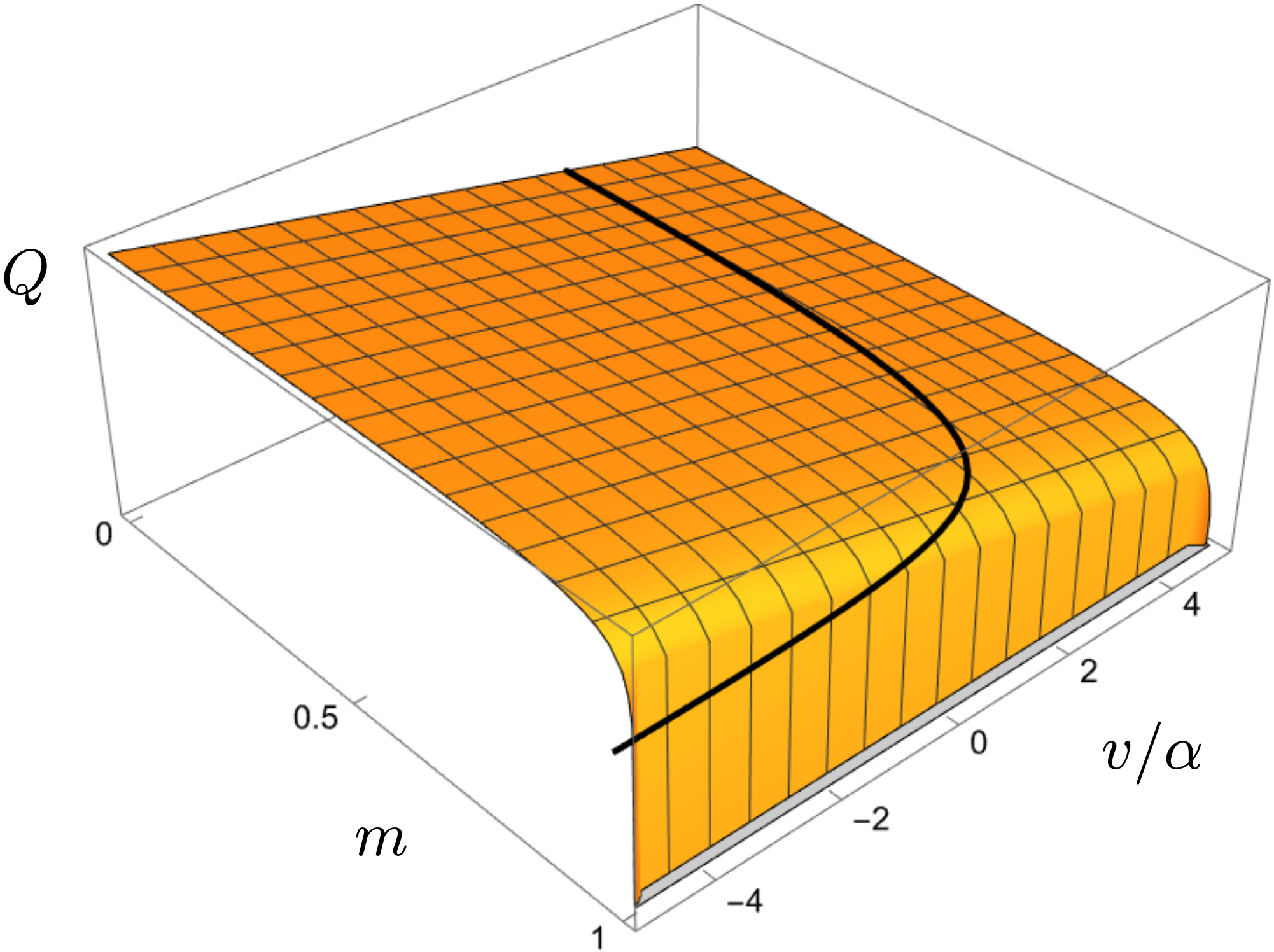}
\caption{The cnoidal average \eqref{CNOREP} as a function of the pointedness $m$ and the velocity $v$, at fixed $\beta/\alpha>0$ and central charge $\nu=1$. The curve where $Q=0$ is shown in black.}
\label{FICNOREP}
\end{figure}

\begin{figure}[t]
\includegraphics[width=.35\textwidth]{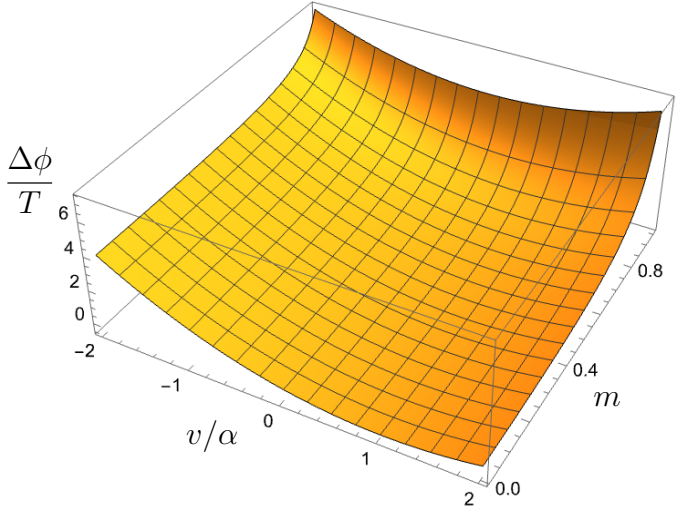}
\caption{The drift velocity \eqref{KEYC}, in arbitrary units, for cnoidal wave solutions of the Korteweg-de Vries equation \eqref{KDV}. Note the highly nontrivial dependence of the drift on the peakedness parameter $m$. Also note the absence of singularity for cnoidal profiles with vanishing average \eqref{CNOREP}, confirming the cancellation in eq.~\eqref{MEGADETH}.}
\label{FIDRIFT}
\end{figure}

Let us focus on cnoidal waves \eqref{CNO}. Since these are travelling waves, the phase shift $\Delta\phi$ is given by \eqref{KEYT}, which involves the following integrals of the wave profile. First, the overall charge \eqref{e4} of the profile is
\be
\label{CNOREP}
Q
=
\frac{4\beta}{\alpha}\frac{K(m)^2}{\pi^2}\Big(m-2+3\frac{E(m)}{K(m)}\Big)-\frac{v-\nu \omega}{\nu \alpha},
\ee
where $E(m)$ is a complete elliptic integral of the second kind. This is plotted in fig.~\ref{FICNOREP} as a function of $m,v$. Note that $Q$ may vanish for any value of the peakedness $m$. Second, the average of $\rho^2$, needed for the Berry phase in \eqref{KEYT}, is given by
\be
\begin{split}
\oint\!\dd x\big(\rho^2 - Q^2\big)
=
96\frac{\beta^2}{\alpha^2}\frac{K(m)^2}{\pi^3}\Big((m-1)K(m)^2\\
- 2(m-2)K(m)E(m)-3E(m)^2\Big).
\end{split}
\ee
The integrand \eqref{DYNI} of the dynamical phase can similarly be expressed in terms of elliptic integrals, although its lengthy formula is omitted here. Remarkably, the total phase shift \eqref{KEYT} for cnoidal waves has the much simpler expression
\be
\label{KEYC}
\begin{split}
\frac{\Delta\phi}{T}
=
\frac{v^2{-}\nu^2\omega^2}{2\alpha\nu^2}-\frac{4\beta}{\alpha\nu\pi^2}K(m)\Big(3 E(m)+(m{-}2)K(m)\Big)\\
+\frac{8\beta^2}{\alpha}\frac{K(m)^4}{\pi^4}.
\end{split}
\ee
This is plotted in fig.~\ref{FIDRIFT} in terms of the cnoidal parameters $m,v$, on which $\Delta\phi$ depends smoothly. In addition, $\Delta\phi$ is nonzero for cnoidal waves with vanishing average, in which case $Q=0$ but eq.~\eqref{KEYC} has a smooth limiting value
\be
\label{MEGADETH}
\begin{split}
\frac{\Delta\phi}{T}
=
\frac{24 \beta^2}{\alpha \pi^4} \; K(m)^2 \Big((m-1) K(m)^2-3 E(m)^2\\
- 2 (m-2) E(m) K(m) \Big).
\end{split}
\ee
Note that this limit is finite even though the dynamical phase and Berry phase in \eqref{KEYT} separately diverge as $Q\to0$. All this can of course be stated without reference to Berry phases, through the general formula \eqref{s135} and its special value \eqref{DRIFT} for travelling waves, upon using the same functional derivative as in eq.~\eqref{DYNI}.

It is instructive to compare the result \eqref{KEYC} with a similar phase shift for cnoidal waves, found in \cite{Oblak:2020jek} upon viewing Korteweg-de Vries dynamics \eqref{KDV} as a geodesic equation on the Virasoro group. In that case, the shift was interpreted as a drift velocity for fluid parcels and could similarly be written as a sum of phases---a dynamical one, a Berry phase and an `anomalous' phase. The relevant orbit representative $Q$ there was not given by the average \eqref{CNOREP}, but involved instead a more complicated argument, and it was similarly found that all three phases diverge at $Q=0$ even though their sum remains finite. However, a key difference with respect to the current setup was the appearance of `orbital bifurcations' occurring at values of $(m,v)$ where the topology of the wave's orbit changes \cite{Oblak:2020myi}. There are no such bifurcations in the present case because \i{any} orbit of the affine U(1) group consists of functions with a fixed average, and is ultimately homotopic to a point. By contrast, the topology of Virasoro orbits is much more intricate (see \eg \cite[chap.~7]{Oblak:2016eij}).

\section{Conclusion}
\label{seconc}

This work was devoted to a natural observable in the semiclassical bosonization of chiral fermions. Namely, we showed that the bosonic field $\phi(x,t)$ picks up a net shift $\Delta\phi$ when the underlying density $\rho(x,t)$ is periodic in time. The phase shift is given by eq.~\eqref{s135} in full generality, and it can be decomposed---for waves with nonzero charge \eqref{e4}---as a sum \eqref{KEY} of a dynamical phase and a Berry phase. In particular, the Berry phase probes the symplectic structure of a coadjoint orbit of the affine U(1) group. The shift $\Delta\phi$ may thus be seen as a probe of the infinite-dimensional geometry of the space of wave profiles, much like the hydrodynamical drift velocity previously discussed in \cite{Oblak:2020jek}.

Such geometric interpretations ultimately stem from Lie-Poisson equations, which are ubiquitous in semiclassical field theory \cite{Khesin}. Our main technical tool here was the fact that semiclassical bosonization \cite{Karabali:1991hm,Iso:1992aa,Das:1991qb,Das:1991uta,Sakita:1996ne,Delacretaz:2022ocm} may be seen as a Lie-Poisson equation on the affine U(1) group, with the phase $\phi(x,t)$ and the density $\rho(x,t)$ playing the role of conjugate variables in the group's cotangent bundle.

While the shift $\Delta\phi$ is indeed a natural observable of bosonization, the question of \i{how} exactly it is meant to be observed requires separate attention. A key consequence of semiclassical bosonization is that shifting $\phi(x,t)$ by a constant $\Delta\phi$ amounts to rotating the fermionic field $\psi(x,t)$ by an overall U(1) phase proportional to $\Delta\phi$. In this sense, one expects $\Delta\phi$ to contribute to the phase picked up by the many-body state of a topological insulator when one of its boundary excitations completes one orbit of its motion. Such phases are crucial \eg in quantum Hall interferometers \cite{Bartolomei:2020qfr,Feldman,Frigerio}, where the $\Delta\phi$ of this work presumably appears as a small correction on top of the Aharonov-Bohm phase that accompanies the motion of an edge state in a magnetic field. One may therefore expect the result \eqref{KEY} to be observable in the future, at least in devices 
sensitive enough to pick up the effects of irrelevant corrections \eqref{e2} to the leading-order gapless theory on the edge. In that respect, cold-atom quantum simulators are perhaps more promising, since they are explicitly sensitive to dispersive or nonlinear perturbations of the conformal field theory on the boundary \cite{Leykam,Leonard:2022ndq,Binanti:2023ozm}.

\section*{Acknowledgements}

We thank Benoit Estienne, Gwendal F\`eve, Nathan Goldman, David Hernandez, Johannes Kellendonk, Gregory Kozyreff, Bastien Lapierre, Leonardo Mazza, Gerbold M\'enard, Per Moosavi, Alberto Nardin and Paul Roux for discussions on related subjects. This work was partly supported by the \i{Partenariat Hubert Curien France/Grèce} project entitled \i{Symétries asymptotiques en gravitation, théories des champs conformes et holographie}.

\appendix

\section{Symplectic geometry on Lie groups}
\label{appa}
\setcounter{equation}{0}
\renewcommand{\theequation}{A.\arabic{equation}}

This appendix collects background material on Lie groups and Lie-Poisson equations, eventually including the effects of central extensions. We refer \eg to \cite{Abraham} or \cite[chap.~5]{Oblak:2016eij} for an introduction to Lie groups and symplectic geometry, and to \cite{Khesin} for Lie-Poisson (or Euler-Arnold) dynamics. The discussion of Berry phases, and the details of their specific appearance in the affine U(1) group, are relegated to appendix \ref{appb}.

\subsection{Lie groups}
\label{APPGROUP}

Here we review the basics of Lie groups, Lie algebras and their duals, and the resulting adjoint and coadjoint representations.

\medskip
\noindent\textbf{Lie groups.}
A \i{Lie group} $G$ is a smooth manifold endowed with a group structure such that multiplication and inversion are smooth maps. We denote its elements as $g$, $h$, etc., the identity element is denoted $\II$, and the inverse of $g$ is written $g^{-1}$. Since multiplication is smooth, fixing any group element $g$ defines two diffeomorphisms $L_g:G\to G:h\mapsto gh$ and $R_g:G\to G:h\mapsto hg$, respectively known as \i{left} and \i{right multiplication} by $g$. Note that $L_g$ and $R_h$ commute, by virtue of associativity, for all $g,h$. \i{Conjugation} by $g$ then is the map $L_g\circ R_{g^{-1}} : h\mapsto ghg^{-1}$.

\medskip
\noindent\textbf{Lie algebra and adjoint representation.}
The \i{Lie algebra} $\mg$ of $G$ is its tangent space at the identity: $\mg\equiv T_{\II}G$. We write its elements as $X$, $Y$, etc.; each such element may be viewed as the `velocity' of any path $\gamma_X(t)$ in $G$ such that $\gamma_X(0)=\II$ and $\der_t\gamma_X(t)\big|_{t=0}=X$ (and similarly for $Y$). The Lie algebra is endowed with a bracket that may be defined in terms of left-invariant vector fields (see \eg \cite[sec.~4.1]{Abraham}), but we will adopt a slightly different (though ultimately equivalent) approach here. Namely, define first the \i{adjoint representation} of $G$ as the map $\Ad:G\to\text{End}(\mg):g\mapsto\Ad_g$ such that
\be
\label{ADAD}
\Ad_g(X)
\equiv
\frac{\der}{\der t}\Big(g\,\gamma_X(t)\,g^{-1}\Big)\Big|_{t=0}.
\ee
In other words, $\Ad_g$ is the differential of conjugation by $g$ at the identity; we wrote this definition above in eq.~\eqref{ADD} for the affine U(1) group. Note that this guarantees that $\Ad_g(X)$ is indeed an element of the Lie algebra (it is the velocity of a path that reaches the identity at $t=0$); one may also verify that $\Ad$ is indeed a representation, \ie that $\Ad_{gh}=\Ad_g\circ\Ad_h$. Now, define the \i{Lie bracket} in $\mg$ by the differential of the adjoint representation,
\be
\label{XYBRA}
[X,Y]
\equiv
\ad_X(Y)
\equiv
\frac{\der}{\der t}\Ad_{\gamma_X(t)}(Y)\Big|_{t=0},
\ee
which is also written in eq.~\eqref{e12} above and defines the adjoint representation of the Lie algebra $\mg$. In what follows, the words `adjoint representation' refer to both $\Ad$ and $\ad$; which of the two is meant will be clear from context.

\medskip
\noindent\textbf{Coadjoint representation.}
Since the Lie algebra $\mg$ is a vector space, it has a dual space $\mg^*$ with elements $p$, $q$, etc. Each such element is a linear map from $\mg$ to $\RR$; we write this map as a pairing
\be
\label{e73}
p:\mg\to\RR:X\mapsto\langle p,X\rangle.
\ee
Such a pairing was introduced above in eq.~\eqref{e14} for the U(1) current algebra. The group $G$ acts on $\mg^*$ according to the dual of the adjoint representation \eqref{ADAD}, \ie the \i{coadjoint representation} $\Ad^*:G\to\text{End}(\mg^*):g\mapsto\Ad^*_g$, given by
\be
\label{e74}
\big<\Ad^*_g(p),X\big>
\equiv
\big<p,\Ad_g^{-1}(X)\big>
\ee
for any $X\in\mg$ and any $p\in\mg^*$. We wrote this in \eqref{COAD} for the U(1) current algebra. Again, one can verify that this is indeed a representation, \ie that $\Ad^*_{gh}=\Ad^*_g\circ\Ad^*_h$. It follows that its differential defines a representation of the Lie algebra: analogously to the adjoint \eqref{XYBRA}, the coadjoint representation of $\mg$ is the map $\ad^*:\mg\to\text{End}(\mg^*):X\mapsto\ad^*_X$ given by
\be
\ad^*_X(p)
\equiv
\frac{\der}{\der t}\Ad^*_{\gamma_X(t)}(p)\Big|_{t=0},
\ee
which is equivalent to
\be
\label{COGEGEN}
\big<\ad^*_X(p),Y\big>
\equiv
-\big<p,[X,Y]\big>
\ee
for any $X,Y\in\mg$ and any $p\in\mg^*$. For the U(1) current algebra, we wrote these definitions in eqs.~\eqref{e17} above. Note that, given any $q\in\mg^*$, we call the set
\be
\cO_q
\equiv
\big\{\Ad^*_gq\,\big|\,g\in G\big\}
\label{s22}
\ee
the \i{coadjoint orbit} of $q$, as in \eqref{s9b} for the affine U(1) group. We also call \i{stabilizer of $q$} the set of group elements $h$ such that $\Ad^*_hq=q$, so the coadjoint orbit \eqref{s22} is diffeomorphic to a quotient space of $G$ by the stabilizer subgroup.

\subsection{Lie-Poisson equations}
\label{appa2}

Let us now briefly review the construction of Lie-Poisson equations on the cotangent bundle of a Lie group. Such systems also go by the name of Euler-Arnold equations, sometimes with the name of Poincar\'e thrown in as well \cite{Khesin,Modin,Holm,Marsden,Marsden90}. The terminology chosen here \cite{Marsden} emphasizes the nature of the group-theoretic and geometric tools involved, while avoiding the restriction to geodesics generally understood in Euler-Arnold systems \cite{Khesin}.

\medskip
\noindent\textbf{Lie groups as configuration spaces.}
Given any manifold $\cM$, its cotangent bundle $T^*\cM$ is a symplectic manifold with a globally exact symplectic two-form $\Omega=-\dd\cA$, where $\cA$ is the Liouville one-form. The simplest and most familiar example is the 2D phase space $\RR^2=T^*\RR$, with coordinates $(q,p)$ and symplectic form $\Omega=\dd q\wedge\dd p=-\dd(p\,\dd q)$. This applies in particular to any Lie group $G$, whose cotangent bundle $T^*G$ may be seen as a phase space. The fact that $G$ is a \i{group} actually makes the bundle structure trivial: one can prove that $T^*G$ is a trivial bundle, equivalent (as a bundle) to the product $G\times\mg^*$, where $G$ is now seen as a configuration space while the dual $\mg^*$ is seen as momentum space; see \eg \cite[app.~A]{Oblak:2020jek}. The symplectic form on $G\times\mg^*$ reads \cite[eq.~(16)]{Oblak:2020jek}
\be
\label{s215t}
\Omega=-\dd\cA%
~~\text{with}~~%
\cA_{(g,p)}
\equiv
\Big(%
\langle p,\dd g\,g^{-1}\rangle,0%
\Big)
\in T_g^*G\oplus T_p^*\mg^*
\ee
where $\dd g\,g^{-1}$ is the right Maurer-Cartan form on $G$. The latter is, by definition, the $\mg$-valued one-form on $G$ such that
\be
\label{e79}
\big(\dd g\,g^{-1}\big)(v)
\equiv
\frac{\der}{\der\tau}\Big|_{\tau=t}\big(g(\tau)\cdot g(t)^{-1}\big)
\equiv
\der_tg\,g^{-1}
\ee
for any tangent vector $v=\der_tg(t)\in T_{g(t)}G$. We will see below that the one-form \eqref{s215t} is essentially a Berry connection and that the right logarithmic derivative \eqref{e79} is the `velocity' of $g(t)$ on the group manifold.

A point in phase space is thus a pair $(g,p)\in G\times\mg^*$, and any choice of Hamiltonian $ H(g,p)$ gives rise to classical equations of motion that determine possible paths $\big(g(t),p(t)\big)$, seen as time-dependent `positions' and `momenta' of the system. For example, when $G=\RR$ is the translation group, the one-form \eqref{s215t} is $\cA=p\,\dd q$ and the logarithmic derivative \eqref{e79} is just $\der_tq(t)$, so the resulting equations of motion are the usual ones of Hamiltonian mechanics.

\medskip
\noindent\textbf{Kinetic Hamiltonians and Lie-Poisson dynamics.}
From now on, let the Hamiltonian \i{only} depend on momenta, so that $ H= H(p)$; this is a setup where energy is exclusively kinetic, without potential. We refer to the ensuing equations of motion as \i{Lie-Poisson equations}, and they turn out to read \cite[app.~B]{Oblak:2020jek}
\begin{align}
\label{EPAGAIN}
\der_tp
&=
-\text{ad}^*_{\dd  H_{p(t)}}\big(p(t)\big),\\
\label{EGAGAIN}
\der_tg\,g^{-1}
&=
-\dd  H_{p(t)}.
\end{align}
Several comments are in order. First, consider the quantities involved: given the Hamiltonian $ H(p)$, its differential $\dd  H_p$ at $p\in\mg^*$ is a linear form from $\mg^*$ to $\RR$, and may thus be seen as a Lie algebra element (any vector space $V$ is isomorphic to its bidual $(V^*)^*\cong V$). It is thus legal to plug $\dd  H_p$ as an argument in the coadjoint representation \eqref{COGEGEN}, which governs the time evolution of the momentum \eqref{EPAGAIN}. The same Lie algebra element plays the role of `velocity' through the equation of motion \eqref{EGAGAIN} for position, where $\der_tg\,g^{-1}$ is the right logarithmic derivative \eqref{e79}.

The structure of eqs.~\eqref{EPAGAIN}--\eqref{EGAGAIN} is hierarchical: the main dynamics lies in the first-order momentum equation of motion \eqref{EPAGAIN}. The latter is nonlinear, unless $ H(p)$ is quadratic (which is exactly what happens in Euler-Arnold equations \cite{Khesin}, as opposed to more general Lie-Poisson equations). Assuming one has solved \eqref{EPAGAIN} with some initial condition $p(0)$, one now has a path $p(t)$ in momentum space. The next step is to integrate eq.~\eqref{EGAGAIN} to deduce $g(t)$ from its velocity \eqref{e79}. We sometimes refer to eq.~\eqref{EPAGAIN} alone as a `Lie-Poisson equation', while eq.~\eqref{EGAGAIN} is called a `reconstruction equation' since it allows one to reconstruct the motion $g(t)$ from the knowledge of $p(t)$; see fig.~\ref{fix}.

One of our key statements above is that the bosonized equations of motion \eqref{e18a}--\eqref{e18b} are reconstructed Lie-Poisson equations for the affine U(1) group. More generally, equations of the form \eqref{EPAGAIN}--\eqref{EGAGAIN} are ubiquitous in nature. This is especially true with quadratic $ H(p)$, in which case the relation \eqref{EGAGAIN} between velocity and momentum is linear, and eqs.~\eqref{EPAGAIN}--\eqref{EGAGAIN} describe geodesics in $G$ with respect to an invariant metric \cite{Khesin}. Here are a few examples:
\begin{itemize}\setlength{\itemsep}{0em}
\item The motion of (freely falling) spinning tops is a Lie-Poisson equation on the rotation group (with a quadratic Hamiltonian).
\item Fluid flows are Lie-Poisson equations on diffeomorphism groups \cite{Khesin,Modin}, notably including the historical case of ideal fluids and the corresponding volume-preserving diffeomorphisms \cite{ArnoldOrigin} (again with a quadratic Hamiltonian).
\item The optimization of unitary quantum gates involves geodesics on the unitary group, hence Lie-Poisson dynamics (again with quadratic $ H$) \cite{Nielsen:2005mkt,Nielsen:2006cea}.
\item The dynamics of Fermi liquids may be seen as a Lie-Poisson equation on the group of canonical transformations of phase space \cite{Delacretaz:2022ocm}. The 1D chiral fermions discussed here are actually a special case of this statement. Indeed, the key role of canonical transformations in bosonization has been known for a while \cite{Karabali:1991hm,Iso:1992aa,Das:1991qb,Das:1991uta,Sakita:1996ne}, and it ultimately leads to Lie-Poisson dynamics on an affine U(1) group through gauge-fixing.
\end{itemize}
For future reference, note a geometric consequence of the Lie-Poisson equation \eqref{EPAGAIN}: it guarantees that $p(t)$ lies in a single coadjoint orbit \eqref{s22}, namely that of the initial momentum $p(0)$. One can therefore write the solution of \eqref{EPAGAIN} as
\be
p(t)
=
\Ad^*_{g(t)}q
\label{s22b}
\ee
where $q$ is such that $p(0)=\Ad^*_{g(0)}q$ and $g(t)$ solves the reconstruction equation \eqref{EGAGAIN}. We emphasize that knowing $q$ and $p(t)$ in eq.~\eqref{s22b} does \i{not} fix $g(t)$ uniquely, since one can always multiply $g(t)$ from the right by a time-dependent stabilizer $h(t)$ of $q$ without affecting eq.~\eqref{s22b}. This will be useful for the discussion of Berry phases in appendix \ref{appb}.

\medskip
\noindent\textbf{Reduction to a coadjoint orbit.}
The fact that the Lie-Poisson momentum \eqref{s22b} lies in a single coadjoint orbit is part of a broader relation between the cotangent bundle $T^*G\cong G\times\mg^*$ and coadjoint orbits of $G$. For the affine U(1) group, this reduction from bundle to orbit was described in section \ref{SEBO}, where it led to a U(1) current algebra \eqref{e19} through a Dirac bracket. Let us therefore sketch, in general terms, how the symplectic form of $G\times\mg^*$ induces a Kirillov-Kostant symplectic form on each coadjoint orbit of $G$ \cite{KirillovLectures}, and how this reduction leads to the aforementioned Dirac bracket \cite{MarsdenRatiu}.

Given a coadjoint orbit \eqref{s22} of $G$ and a point $p=\Ad^*_gq$ thereon, any vector $v$ tangent to $\cO_q$ at $p$ can be written as $v=\ad^*_Xp$ for some Lie algebra element $X\in\mg$. One then defines the \i{Kirillov-Kostant(-Souriau)} symplectic form $\omega$ on $\cO_q$ by
\be
\omega_p(\ad^*_Xp,\ad^*_Yp)
\equiv
\langle p,[X,Y]\rangle\text{ for all }p\in\cO_Q,\,X,Y\in\mg.
\label{ss225}
\ee
The latter is obtained by a so-called symplectic reduction from $T^*G\cong G\times\mg^*$, whose symplectic form is given by \eqref{s215t}. A quick way to establish this link is to note the similarity between \eqref{s215t} and \eqref{ss225}: replacing $p$ by $\Ad^*_gq$ in the one-form $\cA$ of \eqref{s215t}, and considering $q$ as fixed, one obtains the one-form
\be
\label{tilda}
\tilde\cA_g=\langle q,g^{-1}\dd g\rangle\in T^*_gG
\ee
where $g^{-1}\dd g=\Ad_{g^{-1}}(\dd g\,g^{-1})$ is the left Maurer-Cartan form on $G$. (The left Maurer-Cartan form is defined similarly to the right one in \eqref{e79}, except that the multiplication by $g^{-1}$ is from the left, not from the right.) Now taking the exterior derivative of \eqref{tilda} yields a two-form $-\dd\tilde\cA=\langle q,[g^{-1}\dd g,g^{-1}\dd g]\rangle$, which turns out to be the pullback $\pi^*\omega$ of the Kirillov-Kostant symplectic form \eqref{ss225} by the projection $\pi:G\to\cO_q:g\mapsto\Ad^*_gq$ \cite[sec.~5.3.2]{Oblak:2016eij}. In fact, this is why the symplectic flux on the left-hand side of eq.~\eqref{t115} can be evaluated with the unreduced symplectic form \eqref{s215t}, even though it ultimately measures a flux on a coadjoint orbit. A more precise statement can be established thanks to the momentum map $\mu(g,p)\equiv\Ad^*_{g^{-1}}p$ and the Marsden-Weinstein theorem \cite{Marsden:1974dsb}, but we omit this derivation here for brevity.

The reduction from $T^*G\cong G\times\mg^*$ to $\cO_q$ can also be phrased in terms of Poisson and Dirac brackets rather than symplectic forms: see \eg the examples in \cite{Bajnok:2000uv,Mauro:2004cg,Andrzejewski:2020qxt}. The basic idea is that any symplectic form $\omega$ on some manifold $\cM$ defines a bracket of functions on $\cM$, and that the pullback relating the symplectic forms \eqref{s215t} and \eqref{ss225} induces a relation between the Poisson bracket on $T^*G\cong G\times\mg^*$ and the Dirac bracket on $\cO_q$. Here we skip this argument for brevity and refer instead to \cite{MarsdenRatiu} for a general derivation and many more references on the subject.

\subsection{Adding central extensions}

Affine Lie groups are centrally extended, so it is crucial for our purposes to revisit the above in that context. Let therefore $G$ be a Lie group, and let $\sfC:G\times G\to\RR$ be a smooth map such that $\sfC(g,\II)=\sfC(\II,g)=0$ and
\be
\sfC(f,g)+\sfC(fg,h)
=
\sfC(g,h)+\sfC(f,gh).
\label{t225}
\ee
Then $\sfC$ is known as a (real) \i{two-cocycle} on $G$, and it defines a \i{central extension} $\widehat G=G\times\RR$ whose elements are pairs $(g,a)$, $(h,b)$, etc.\ with a group operation
\be
(g,a)\cdot(h,b)
\equiv
\big(gh,a+b+\sfC(g,h)\big).
\label{s225}
\ee
The latter is associative by virtue of the cocycle condition \eqref{t225}. The identity in $\widehat G$ is $(\II,0)$ and the inverse is given by $(g,a)^{-1}=(g^{-1},-a-\sfC(g,g^{-1}))$. In practice, one can always choose the cocycle to be such that $\sfC(g,g^{-1})=0$ for all $g$, so that the inverse is just $(g,a)^{-1}=(g^{-1},-a)$.\footnote{Indeed, let $\sfK:G\to\RR$ be a map, define $\widetilde\sfC(g,h)\equiv\sfC(g,h)+\sfK(gh)-\sfK(g)-\sfK(h)$, and let $\widetilde G=G\times\RR$ be the extended group whose multiplication is \eqref{s225} with $\sfC$ replaced by $\widetilde\sfC$. Then the map $\alpha:\widehat G\to\widetilde G:(g,a)\mapsto(g,a+\sfK(g))$ is an isomorphism, and choosing $\sfK(g)=\sfK(g^{-1})=\tfrac{1}{2}\sfC(g,g^{-1})$ yields $\widetilde\sfC(g,g^{-1})=0$.} We assume from now on that such a choice has been made. In the main text, a centrally extended group operation \eqref{s225} was used in \eqref{e24} to define the affine U(1) group, with $G=C^{\infty}(S^1)$ and $\sfC(\phi,\chi)=\oint\tfrac{\dd x}{4\pi}\phi\,\der_x\chi$ such that $\sfC(\phi,-\phi)=0$.

\medskip
\noindent\textbf{Adjoint representation and Lie algebra.}
All the definitions given above for adjoint and coadjoint representations carry over to centrally extended groups. Thus, the centrally extended Lie algebra $\widehat\mg=\mg\oplus\RR$ consists of pairs $(X,a)$, $(Y,b)$, etc.\footnote{The same letters $a$, $b$, etc.\ denote central entries in both the group $\widehat G$ and its algebra $\widehat\mg$. This is similar to the abuse of notation in footnote \ref{foot1}, since the subgroup of central entries is a vector group.} It is acted upon by an adjoint action $\widehat\Ad$ that can be related to the adjoint $\Ad$ of $G$ through the basic definition \eqref{ADAD}. Explicitly,
\be
\begin{split}
\widehat\Ad_{(g,a)}(X,b)
&\equiv
\frac{\der}{\der t}\Big|_{t=0}\Big[(g,a)\cdot\big(\gamma_X(t),tb\big)\cdot(g,a)^{-1}\Big]\\
&=
\big(\Ad_gX,b-\langle\sfS[g],X\rangle\big),
\end{split}
\label{s215}
\ee
where $\langle\cdot,\cdot\rangle$ denotes the pairing \eqref{e79} of $\mg^*$ and $\mg$, and we have defined
\be
\big<\sfS[g],X\big>
\equiv
-\frac{\der}{\der t}\Big|_{t=0}
\Big(\sfC\big(g,\gamma_X(t)\big)+\sfC\big(g\gamma_X(t),g^{-1}\big)\Big).
\label{t215}
\ee
Here the map $\sfS:G\to\mg^*:g\mapsto\sfS[g]$ is known as the \i{Souriau cocycle}: it is actually a $\mg^*$-valued one-cocycle on $G$ (since $\sfS[gh]=\Ad^*_{h^{-1}}\sfS[g]+\sfS[h]$ in terms of the coadjoint representation \eqref{e74} of $G$).

Note in \eqref{s215} that the central entry $a$ does not affect the adjoint operator, so we simplify notation and write $\widehat\Ad_{(g,a)}\equiv\widehat\Ad_g$ from now on. This notation was used in eq.~\eqref{ADD} for the adjoint of the affine U(1) group, where the Souriau cocycle is just $\sfS[\phi]=\der_x\phi$. The resulting centrally extended bracket is defined by \eqref{XYBRA}, and is related to the Lie bracket of $\mg$ through
\be
\big[(X,a),(Y,b)\big]
=
\big([X,Y],-\langle\sfs[X],Y\rangle\big)
\label{s205}
\ee
where $\sfs[X]\equiv\der_t\big|_0\sfS[\gamma_X(t)]$ is the infinitesimal Souriau cocycle. For the U(1) current algebra, this bracket was written in \eqref{e12} with $\sfs[\phi]=\der_x\phi$. 

\medskip
\noindent\textbf{Coadjoint representation.}
Let us turn to the coadjoint representation of $\widehat G$, which acts on the dual space $\widehat\mg{}^*=\mg^*\oplus\RR$ that consists of pairs $(p,\nu)$.\footnote{Writing $\nu$ for a central charge is perhaps uncommon, but it is justified here by the fact that the central charge counts the number of edge channels, \ie the filling fraction in the quantum Hall context.} The pairing between $\widehat\mg{}^*$ and $\widehat\mg$ extends \eqref{e73} as $\langle(p,\nu),(X,a)\rangle=\langle p,X\rangle+\nu a$, which was written in eq.~\eqref{e14} for the affine U(1) group. According to the definition \eqref{e74}, the coadjoint action of $\widehat G$ is then related to that of $G$ by
\be
\widehat\Ad{}^*_g(p,\nu)
=
\big(\Ad^*_gp-\nu\,\sfS[g^{-1}],\nu\big)
\label{t205}
\ee
where $\sfS$ is again the Souriau cocycle defined in \eqref{t215}. In the affine U(1) group, this definition was written in eq.~\eqref{COAD}. The corresponding coadjoint representation of the Lie algebra $\widehat\mg$ is
\be
\widehat\ad{}^*_X(p,\nu)
=
\big(\ad^*_Xp+\nu\,\sfs[X],0\big),
\label{tt205}
\ee
which was written in eq.~\eqref{e17} for the affine U(1) group. Note from \eqref{t205} that the central charge $\nu$ is constant in any coadjoint orbit \eqref{s22} of a centrally extended group.

\medskip
\noindent\textbf{Lie-Poisson equations.}
We finally consider centrally extended Lie-Poisson equations. The phase space is thus a product $\widehat G\times\widehat\mg{}^*$, with a symplectic form $\widehat\Omega=-\dd\widehat\cA$ whose Liouville one-form extends the one written for $G\times\mg$ in eq.~\eqref{s215t}. One explicitly finds
\be
\begin{split}
\widehat\cA_{((g,a),(p,k))}
=
\Big(%
\langle p,\dd g\,g^{-1}\rangle
+
\nu\,\dd\sfC_g,\nu\,\dd a\,;0,0
\Big)\\
\in
T_g^*G\oplus\RR^*\oplus T^*_p\mg^*\oplus\RR^*,
\end{split}
\label{s13b}
\ee
where $\dd\sfC$ is the one-form on $G$ defined by $\dd\sfC(\der_tg)\equiv\der_{\tau}\sfC(g(\tau),g(t)^{-1})$. In the affine U(1) group, the Liouville one-form is the Berry connection in the Berry phase \eqref{s11b}. Similarly to the centreless case above, any choice of Hamiltonian on $\widehat G\times\widehat\mg^*$ implies Hamiltonian equations of motion. Lie-Poisson dynamics occurs for a Hamiltonian $H= H(p)$ that only depends on momenta $p\in\mg^*$. Then the extended form of the Lie-Poisson equation of motion \eqref{EPAGAIN} is
\be
(\der_tp,\der_t\nu)
=
-\widehat\ad{}^*_{\dd  H_p}(p,\nu)
\!\stackrel{\text{\eqref{tt205}}}{=}\!
\big({-}\ad^*_{\dd  H_p}p{-}\nu\,\sfs[\dd  H_p],0\big).
\label{s21b}
\ee
This shows that Lie-Poisson dynamics always occurs at constant central charge ($\der_t\nu=0$), and that momentum dynamics is only modified from its centreless form \eqref{EPAGAIN} by the Souriau cocycle defined around \eqref{s205}. In the affine U(1) group, the analogue of eq.~\eqref{s21b} was written in \eqref{ss85}. Note that the \i{entire} dynamics of $\rho$ in that case stems from the central extension and the Souriau cocycle, because the underlying centreless group is Abelian.

As explained in section \ref{SAFFI} and appendix \ref{appa2}, the second step of Lie-Poisson equations is to solve the reconstruction equation \eqref{EGAGAIN}. The latter involves the right logarithmic derivative \eqref{e79} of $g(t)$, which takes a more complicated form in centrally extended groups. Indeed, let $(g(t),a(t))$ be a path in $\widehat G=G\times\RR$ and use the multiplication \eqref{s225} to find
\be
\begin{split}
&\frac{\der}{\der\tau}\Big|_{\tau=t}\Big[\big(g(\tau),a(\tau)\big)\cdot\big(g(t),a(t)\big)^{-1}\Big]\\
&=
\Big(%
\der_tg\,g^{-1},\der_ta+\der_{\tau}\big|_t\sfC\big(g(\tau),g(t)^{-1}\big)
\Big),
\end{split}
\ee
where $\der_tg\,g^{-1}$ is the right logarithmic derivative \eqref{e79} of $g(t)$ alone, and the derivative of the cocycle $\sfC$ is the same as in the one-form \eqref{s13b}. Plugging this in the reconstruction equation \eqref{EGAGAIN} yields
\be
\label{s21bb}
\der_tg\,g^{-1}
=
-\dd  H_{p(t)},
\qquad
\der_ta+\der_\tau\big|_t\sfC\big(g(\tau),g(t)^{-1}\big)=0,
\ee
which was written in \eqref{ss75} for the affine U(1) group. Note that the reconstruction equation for the central entry $a$ is readily solved as
\be
a(t)
=
a(0)
-
\int_0^t\dd s\,\der_{\tau}\big|_{\tau=s}\sfC(g(\tau),g(s)^{-1}).\label{atsol}
\ee
We show in appendix \ref{appb2} that this contributes an `anomalous phase' to the phase shift of $g(t)$ when the solution $p(t)$ of \eqref{s21b} is periodic.

\section{Berry phases in Lie-Poisson equations}
\label{appb}
\setcounter{equation}{0}
\renewcommand{\theequation}{B.\arabic{equation}}

In this appendix, we explain in general terms how periodic solutions of the Lie-Poisson equation \eqref{EPAGAIN} give rise to Berry phases through the reconstruction equation \eqref{EGAGAIN}. The presentation begins with the generic case, then addresses the effects of central extensions. For more on this subject, see \eg \cite{Marsden90,Montgomery}, or the more recent \cite{Oblak:2020jek} where central extensions and the ensuing `anomalous phases' are studied in detail.

\subsection{Lie-Poisson phases in general}

Let $G$ be a Lie group, $ H$ a Hamiltonian on $\mg^*$, and let $(g(t),p(t))\in G\times\mg^*$ be a solution of the Lie-Poisson equations \eqref{EPAGAIN}--\eqref{EGAGAIN}. We assume that $p(t)$ has some period $T$, so that $p(t+T)=p(t)$. The question is: what does this imply for the reconstructed solution $g(t)$ of \eqref{EGAGAIN}, specifically for the difference between $g(T)$ and $g(0)$?

The solution is sketched in fig.~\ref{fi4}, being understood that `densities' $\rho$ are now momenta $p\in\mg^*$, while `phases' $\phi$ are group elements $g\in G$. Indeed, writing the time-dependent momentum as in eq.~\eqref{s22b} and using periodicity implies that $g(0)^{-1}g(T)$ stabilizes $q$. The idea then is to integrate the Berry connection \eqref{s215t} along the closed path obtained by first following the reconstructed motion $g(t)$, then using a path in the stabilizer of $q$ to connect $g(T)$ back to $g(0)$. This integral is the Berry phase picked up by a coherent state acted upon by the time-dependent transformations $g(t)$ \cite{Bose,Boya,Zalesny:2000zz,Oblak:2017ect}. For the affine U(1) group, such an integral `in two pieces' was introduced in \eqref{s115}, and the corresponding integral of $\cA$ was schematically written in \eqref{t115}. The resulting holonomy $\oint\cA$ affects the mismatch between $g(0)$ and $g(T)$, since one can write
\be
\label{e29}
\text{Berry}
\equiv
\oint\cA
\stackrel{\text{\eqref{s215t}}}{=}
\!\!\int\limits_0^T\!\dd t\big< p(t),\der_tg\,g^{-1}\big>+\int\limits_T^{T'}\!\dd t\langle q,h^{-1}\der_th\rangle
\ee
where $T'>T$ is arbitrary and the path $h(t)$ is such that $\Ad^*_{h(t)}q=q$, with $h(T)=\II$ and $h(T')=g(T)^{-1}g(0)$. The reconstruction equation \eqref{EGAGAIN} then gives a rewriting of the first term as an integral of $\langle p,\dd  H_p\rangle$, \ie a dynamical phase. Thus, eq.~\eqref{e29} can be recast as
\be
\label{s21t}
\int_T^{T'}\dd t\,\langle q,h^{-1}\der_th\rangle
=
\underbrace{\oint\cA}_{\ds\text{Berry}}
+
\underbrace{\oint\dd t\langle p,\dd  H_p\rangle.}_{\ds\text{Dynamical}}
\ee
This is a general result that holds for any periodic solution $p(t)$ of the Lie-Poisson equation \eqref{EPAGAIN}. If the Hamiltonian is a homogeneous function of $p$, say with degree $d$, then the dynamical phase is $dET$ where $E= H(p)$ is the conserved energy along $p(t)$.

While the interpretation of the two terms on the right-hand side of \eqref{s21t} is clear, that of the left-hand side is less obvious. It essentially measures the difference between the group elements $g(0)$ and $g(T)$, since the path $h(t)$ was chosen below eq.~\eqref{e29} to connect the identity to $g(T)^{-1}g(0)$. To make this more explicit, let us now assume that the stabilizer of $q$ is Abelian, so that its universal cover is a vector group. Then eq.~\eqref{s21t} can be written as
\be
-\langle q,\Delta v\rangle
=
\oint\cA
+
\oint\dd t\langle p,\dd  H_p\rangle
\label{s21q}
\ee
where $\Delta v\equiv g(0)^{-1}g(T)$ is the vector, in the universal cover of the stabilizer of $q$, that lifts the group element $g(0)^{-1}g(T)$ (with the convention that $\Delta v=0$ is the lift of the identity). In the special case where the stabilizer is a one-dimensional group, $\Delta v$ is just a phase shift $\Delta\phi$. This is essentially the phase shift studied in this work, up to a subtlety involving central extensions that will be addressed shortly.

Eq.~\eqref{s21q} is a universal formula for the `rotation' of a system once its momentum has completed one orbit of its motion. Its best-known application is the rotation of rigid bodies, where the Hamiltonian is quadratic so that the dynamical phase is just a product of period and energy, and the Berry phase coincides (up to normalization) with that of a slowly rotating qubit \cite{Berry,Montgomery}. However, it should be clear from the derivation shown here that the presence of such phases is a much more general phenomenon \cite{Marsden90,Oblak:2020jek}. The case of the affine U(1) group requires one last ingredient, namely a discussion of `anomalous phases' due to the central extension.

\subsection{Lie-Poisson phases for centrally extended groups}
\label{appb2}

Consider a centrally extended group $\widehat G=G\times\RR$ with Lie-Poisson equation \eqref{s21b}, and assume that the momentum $p(t)$ is periodic as above. The argument leading to eq.~\eqref{s21t} remains valid, but the Berry phase now involves the centrally extended connection \eqref{s13b}:
\be
\label{s215b}
\int_T^{T'}\dd t\,\big<(q,\nu),h^{-1}\der_th\big>
=
\oint\widehat\cA
+
\oint\dd t\langle p,\dd  H_p\rangle.
\ee
Note here that the central charge does not contribute to the dynamical phase, although it does contribute to both the Berry phase and the left-hand side. Indeed, the stabilizer of $(q,\nu)$ certainly contains the $\RR$ subgroup of central entries $(0,a)\in\widehat G$, and the difference $\Delta a\equiv a(T)-a(0)$ is generally nonzero owing to the solution \eqref{atsol} of the centrally extended Lie-Poisson equation. This leads to an `anomalous phase' $\nu \Delta a$ on top of those written in \eqref{s21q}: with the same conventions for the universal cover of the stabilizer of $q$, eq.~\eqref{s215b} yields \cite{Oblak:2020jek}
\be
-\langle q,\Delta v\rangle
-\nu\Delta a
+\nu\sfC\big(g(T)^{-1},g(0)\big)
=
\oint\widehat\cA
+
\oint\dd t\langle p,\dd  H_p\rangle.
\ee
Now assuming that the stabilizer of $(q,\nu)$ only consists of U(1) rotations and central elements, and using the solution \eqref{atsol}, one finally finds
\be
\label{fennec}
\begin{split}
&-\langle q,\Delta\phi\rangle
=
\underbrace{\oint\widehat\cA}_{\ds\text{Berry}}
+
\underbrace{\oint\dd t\langle p,\dd  H_p\rangle}_{\ds\text{Dynamical}}\\
&-
\underbrace{\nu\int_0^T\dd t\,\der_{\tau}\big|_t\sfC(g(\tau),g(t)^{-1})
-
\nu\,\sfC\big(g(T)^{-1},g(0)\big).}_{\ds\text{Anomalous}}
\end{split}
\ee
Here $\Delta\phi$ is viewed as a Lie algebra element, upon identifying the universal cover of the stabilizer with its Lie algebra. The value of $\Delta\phi$ is clearly affected by the anomalous phase inherited from the solution \eqref{atsol} \cite{Oblak:2020jek}. As we now explain, the effect of the anomalous phase is remarkably mild in the case of the affine U(1) group.

\medskip
\noindent\textbf{Application to chiral density waves.}
The affine U(1) group was defined in section \ref{SAFFI}. It is the (Heisenberg) central extension of the vector group $C^{\infty}(S^1)$, with a cocycle $\sfC(\phi,\chi)=\oint\tfrac{\dd x}{4\pi}\phi\,\der_x\chi$ visible \eg in the group operation \eqref{e24}. Given a periodic solution $\rho(\cdot,t)$ of the Lie-Poisson equation \eqref{ss85}, the data needed to write down eq.~\eqref{fennec} is as follows:
\begin{itemize}
\item The pairing $\langle q,\Delta\phi\rangle=Q\,\Delta\phi$ boils down to the product of the total charge \eqref{e4} with the U(1) phase shift $\Delta\phi$, which one seeks to evaluate.
\item The dynamical phase in \eqref{fennec} involves the (functional) differential of the bosonic Hamiltonian:
\be
\label{er1}
\oint\dd t\langle p,\dd  H_p\rangle
\stackrel{\text{\eqref{e14}}}{=}
\oint\frac{\dd t\,\dd x}{2\pi}\rho(x,t)\frac{\delta  H}{\delta\rho(x,t)}.
\ee
This is the same dynamical phase as in eq.~\eqref{KEY}.
\item The Berry phase in \eqref{fennec} is a holonomy of the centrally extended connection \eqref{s13b}, and was written in eq.~\eqref{s11b} for the affine U(1) group:
\begin{align}
\oint\widehat\cA
&\stackrel{\text{\textcolor{white}{(A.16)}}}{=}
\Big<(Q,\nu),\der_{\tau}\Big|_{t}\big({-}\phi(\cdot,t),0\big)\cdot\big(\phi(\cdot,\tau),0\big)\Big>\\
&\stackrel{\text{\eqref{s225}}}{=}
-\nu\,\der_{\tau}\Big|_{t}\sfC\big(\phi(\cdot,t),\phi(\cdot,\tau)\big)\\
&\stackrel{\text{\eqref{e24}}}{=}
\nu\int\frac{\dd t\,\dd x}{4\pi}\der_t\phi\,\der_x\phi.
\label{tt28}
\end{align}
\item Finally, in the affine U(1) case, the term $\sfC(g(T)^{-1},g(0))$ in eq.~\eqref{fennec} vanishes owing to the form of the cocycle. The entire anomalous phase in \eqref{fennec} thus reads
\be
\label{er3}
\begin{split}
-\nu\int_0^T\dd t\,\der_{\tau}\big|_t\sfC(g(\tau),g(t)^{-1})\\
=
\nu\int\frac{\dd t\,\dd x}{4\pi}\der_t\phi\,\der_x\phi,
\end{split}
\ee
which coincides with the Berry phase \eqref{tt28}.
\end{itemize}
It is now a simple matter to write down the full phase shift \eqref{fennec} for the affine U(1) group: putting together eqs.~\eqref{er1}--\eqref{er3} reproduces eq.~\eqref{KEY} for $\Delta\phi$, as desired. Note that this involves a sum of the identical phases \eqref{tt28} and \eqref{er3}, leading to a factor two mismatch between the Berry phase \eqref{s11b} and the one that appears in eq.~\eqref{KEY}.

\addcontentsline{toc}{section}{References}

\end{document}